\documentclass[a4paper]{JHEP3}
\usepackage{amsmath,latexsym,amssymb}
\usepackage[latin1]{inputenc}
\usepackage{rotating}
\usepackage{Macros}

\usepackage{framed}
\usepackage{comment}

\title{Electric/magnetic deformations of \boldmath $S^3$ and
  $\mathrm{AdS}_3$ \unboldmath, and geometric cosets\thanks{ Research
    partially supported by the EEC under the contracts HPRN-CT-2000-00122,
    HPRN-CT-2000-00131, HPRN-CT-2000-00148.} }

\author{ Dan Israël$^\Diamond$, Costas Kounnas$^\Diamond$, Domenico
  Orlando$^\spadesuit$, P. Marios Petropoulos$^\spadesuit$\\

${}^{\phantom{\spadesuit}\Diamond}$
\begin{minipage}[t]{.7\textwidth}
  Laboratoire de Physique Théorique de l'Ecole Normale
  Sup{é}rieure\footnote{Unité mixte du CNRS et de l'Ecole Normale
    Sup{é}rieure, UMR 8549.} \\
  24 rue Lhomond, 75231 Paris Cedex 05, France
\end{minipage}

${}^{\phantom{\Diamond}\spadesuit}$
\begin{minipage}[t]{.7\textwidth}
  Centre de Physique Théorique, Ecole Polytechnique\footnote{Unité
    mixte du CNRS et de l'Ecole Polytechnique, UMR 7644.} \\
  91128 Palaiseau, France
\end{minipage}

\bigskip

E-mail: \email{israel@lpt.ens.fr}, \email{kounnas@lpt.ens.fr},
\email{orlando@cpht.polytechnique.fr}, \email{marios@cpht.polytechnique.fr}
}

\abstract{We analyze asymmetric marginal deformations of $SU(2)_k$ and
  $SL(2,\mathbb{R})_k$ \textsc{wzw} models. These appear in heterotic string
  backgrounds with non-vanishing Neveu--Schwarz three-forms plus electric or
  magnetic fields, depending on whether the deformation is elliptic,
  hyperbolic or parabolic. Asymmetric deformations create new families of
  exact string vacua. The geometries which are generated in this way,
  deformed $S^3$ or $\mathrm{AdS}_3$, include in particular
  \emph{geometric cosets} such as $S^2$, $\mathrm{AdS}_2$ or $H_2$. Hence,
  the latter are consistent, exact conformal sigma models, with electric or
  magnetic backgrounds. We discuss various geometric and symmetry properties
  of the deformations at hand as well as their spectra and partition
  functions, with special attention to the supersymmetric $\mathrm{AdS}_2 ×
  S^2$ background. We also comment on potential holographic applications.}
  \preprint{
  LPTENS-04/26 \\
  CPTH-RR014.0404 \\
  hep-th/0405213 \\}

\begin{document}

\setcounter{footnote}{0}
\renewcommand{\thefootnote}{\arabic{footnote}}
\setcounter{section}{0}

\section{Introduction}

Near-horizon geometries of
NS5-branes~\cite{Kounnas:1990ud,Callan:1991dj,Antoniadis:1994sr}, NS5/F1 or
S-dual versions of those~\cite{Antoniadis:1990mn,Boonstra:1998yu} have been
thoroughly analyzed over the past years. These involve $\mathrm{AdS}_3$ or
$S^3$ spaces and turn out to be exact string backgrounds, tractable beyond
the supergravity approximation. They offer a unique setting in which to
analyze AdS/\textsc{cft} correspondence, black-hole physics, little-string
theory\dots

An important, and not yet unravelled aspect of such configurations is the
investigation of their moduli space. String propagation in the above
backgrounds is described in terms of some exact two-dimensional conformal
field theory.  Hence, marginal deformations of the latter provide the
appropriate tool for exploring the moduli of the corresponding string vacua.

A well-known class of marginal deformations for \textsc{wzw} models are
those driven by left-right current
bilinears~\cite{Chaudhuri:1989qb,Forste:2003km}: $\int\di^2 z \ J \bar J$.
These can be ``symmetric'' in the sense that both $J$ and $\bar J$ are
generators of the affine algebra of the \textsc{wzw} model.  However, asymmetric
deformations can also be considered, where either $J$ or $\bar J$ correspond
to some other $U(1)$, living outside of the chiral algebra of the \textsc{wzw}
model. These deformations describe the response of the system to a finite (chromo)
electric or magnetic background field and since these deformations are
exactly marginal, the gravitational back-reaction is properly taken into
account at any magnitude of the external
field~\cite{Kiritsis:1995ta,Kiritsis:1995iu}.

The purpose of this note is to report on the asymmetric deformations of the
$SL(2, \mathbb{R})_k$ heterotic string background. Since $SL(2, \mathbb{R})$
has time-like, null and space-like generators, three distinct asymmetric
deformations are possible, corresponding respectively to magnetic,
electromagnetic and electric field backgrounds. The former case has been
recently analyzed from this perspective~\cite{Israel:2003cx} and the
corresponding deformation was shown to include G\"odel space--time, which is
therefore promoted to an exact string background (despite the caveats of
closed time-like curves). The latter case, on the other hand, corresponds to
a new deformation, which connects $\mathrm{AdS}_3$ with $\mathbb{R} \times
\mathrm{AdS}_2$.

This observation is far reaching: while symmetric deformations usually
connect \textsc{wzw} models to some $U(1)$-gauged version of
them~\cite{Giveon:1994ph}, asymmetric deformations turn out to connect the
original theory to some \emph{geometric coset,} with \emph{electric} or
\emph{magnetic} background fields. This holds for the $SU(2)_k$, where the
limiting magnetic deformation has $\mathbb{R} \times S^2$ geometry, and can be
generalized to any \textsc{wzw} model: geometric cosets with electric or
magnetic background fields provide thus \emph{exact} string vacua. Here we
focus on the $S^2$ and AdS$_2$ examples, previously discussed as heterotic
coset constructions in~\cite{Johnson:1995kv} (see also \cite{Lowe:1994gt}).
Moreover, they both enter in the near-horizon geometry of the
four-dimensional Reissner--Nordstr\"om extremal black hole, AdS$_2 \times S^2$,
which is here shown to be an exact string vacuum. We also show how $H_2$
appears as an exact \textsc{cft}, although this background is of limited
interest for string theory because of lack of unitarity.

The paper is organized as follows. First we review the magnetic deformation
of $S^3$, appearing in the framework of the $SU(2)_k$ \textsc{wzw} model.
The appearance of the two-sphere plus magnetic field as exact string
background is described in Sec.~\ref{sphere}, where we also determine the
corresponding partition function. The $\mathrm{AdS}_3$ case is analyzed in
Sec.~\ref{ads}, where its asymmetric deformations are described in detail
from geometrical and two-dimensional-\textsc{cft} points of view. We also
investigate their spectra. Limiting deformations are discussed in
Sec.~\ref{gencos}. There, we show how to reach the $\mathrm{AdS}_2=SL(2,
\mathbb{R})/U(1)$ geometric coset with electric field. These backgrounds are
consistent and exact string vacua.

The $H_2$ geometric coset of AdS$_3$ is also shown to appear on the line of
magnetic deformation, with imaginary magnetic field though. The near-horizon
geometry of the four-dimensional Reissner--Nordstr\"om extremal black hole is
further discussed in Sec.~\ref{ads2s2}. Section \ref{out} contains a
collection of final comments, where we sort various geometries that should
be investigated in order to get a comprehensive picture of the general
AdS$_3$ landscape, and its connection to other three-dimensional geometries.
Four appendices provide some complementary/technical support. Appendix
\ref{def} sets the general framework for geometric deformations of a metric,
designed to keep part of its original isometry. Appendices \ref{sph} and
\ref{antids} contain material about $SU(2)$ and $SL(2,\mathbb{R})$ groups 
manifolds. A reminder of low-energy field equations for the bosonic degrees of freedom of
heterotic string is given in App.~\ref{eom}.

\section{Magnetic deformation of \boldmath $S^3$ \unboldmath}\label{sphere}

$SU(2)_k$ \textsc{wzw} model magnetic deformations were analyzed
in~\cite{Kiritsis:1995ta,Kiritsis:1995iu}, 
for both type II and heterotic string
backgrounds. For concreteness, we will concentrate here on the latter case.
In contrast to what happens in flat space--time, these deformations are
truly marginal in the background of a three-sphere plus \textsc{ns} flux,
and preserve $N = (1,0)$ world-sheet supersymmetry.

Consider heterotic string on $\mathbb{R}^{1,3} × S^3 × T^4$. The theory is
critical provided we have a linear dilaton living on the
$\mathbb{R}^{1,3}$, with background charge $Q=1/\sqrt{k+2}$, where $k$ is
the level of the bosonic $SU(2)_{\rm L} × SU(2)_{\rm R}$ affine algebra. The
target-space geometry is the near-horizon limit of the solitonic
NS5-brane~\cite{Kounnas:1990ud,Callan:1991dj,Antoniadis:1994sr}.

The two-dimensional $N=(1,0)$ world-sheet action corresponding to the $S^3$
factor is
\begin{equation}
  S_{SU(2)_k} =\frac{1}{2\pi} \int {\rm d}^2 z  \left\{\frac{k}{4}
    \left( \d \alpha \db \alpha + \d \beta \db \beta + \d \gamma \db
      \gamma + 2 \cos \beta \, \d \alpha \db \gamma \right) +
    \sum_{a=1}^{3} \psi^a \db \psi^a\right\}, \label{SU2WZW}
\end{equation}
where $\psi^a$ are the left-moving free fermions, superpartners of the bosonic
$SU(2)_k$ currents, and $(\alpha,\beta,\gamma)$ are the usual Euler angles
parameterizing the $SU(2)$ group manifold (see App.~\ref{sph} for a
reminder). In this parameterization, the chiral currents of the Cartan
subalgebra read:
\begin{equation}
  J^3=k\left(\d \gamma +\cos \beta \, \partial \alpha \right)\ , \ \
  \bar J^3=k\left(\db \alpha +\cos \beta \, \db \gamma \right)
\end{equation}
(Tab.~\ref{tab:su2-currents}) with the following short-distance expansion:
\begin{equation}
  J^3(z) J^3(0)={k \over 2z^2} + {\rm reg.}
\end{equation}
and similarly for the right-moving one. The left-moving fermions transform
in the adjoint of $SU(2)$. There are no right-moving superpartners but a
right-moving current algebra with total central charge $c=16$ (realized
\emph{e.g.} in terms of right-moving free fermions). The currents of the
latter are normalized so that the Cartan generators $\bar J^i_G$ of the
group factor $G$ satisfy the following short-distance expansion:
\begin{equation}
  \bar J^i_G(z) \bar J^j_G(0)={k_G h^{ij} \over 2 \bar z^2} + {\rm
    reg.} \ , \ \ i, j =1, \ldots, {\rm rank}(G)
\end{equation}
with $h^{ij}{\vphantom k}= f^{ik}_{\hphantom{ik}\ell}\, f^{\ell
  j}_{\hphantom{ij}k}\big/g^\ast$, $f^{ij}_{\hphantom{ik}k}$ and $g^\ast$ being
the structure constants and dual Coxeter number of the group $G$.

The background metric and \textsc{NS} two-form are read off directly from
(\ref{SU2WZW}):
\begin{align}
  \di s^2 &= \frac{k}{4}\left[\di \beta^2 + \sin^2 \beta \di \alpha^2 +\left(\di \gamma +
      \cos
      \beta \di \alpha \right)^2 \right]\label{S3met}, \\
  B &= \frac{k}{4} \cos \beta \di \alpha \land \di \gamma.
\label{S3ant}
\end{align}
They describe a three-sphere of radius $L=\sqrt{k}$ and a \textsc{ns}
three-form whose field strength is $\di B=\frac{2}{\sqrt{k}}\, \omega_{[3]}$
($\omega_{[3]}$ stands for the volume form given in Eq.~(\ref{eq:su2-vf})).

A comment is in order here. In general, the background fields $G_{ab}$, $B_
{ab}$, \dots receive quantum corrections due to two-dimensional
renormalization effects controlled by $\alpha'$ (here set equal to one).  This
holds even when the world-sheet theory is exact. Conformal invariance
requires indeed the background fields to solve Eqs. (\ref{beta}) that
receive higher-order $\alpha'$ corrections. The case of \textsc{wzw} models is
peculiar in the sense that the underlying symmetry protects $G_{ab}$ and $B_
{ab}$ from most corrections; these eventually boil down to the substitution
$k \to k+2$ in Eqs.  (\ref{S3met}) and (\ref{S3ant}), see
\emph{e.g.}~\cite{Tseytlin:1994my}.

Notice finally that the $SU(2)_k$ plus linear dilaton background introduces
a mass gap with respect to flat space: $\mu^2 = {1/ (k+2)}$. This plays the
role of infra-red regulator, consistent with all string requirements
including worldsheet supersymmetry.

\subsection{Squashing the three-sphere}
\label{sec:squashing-sphere}

We now turn to the issue of conformal deformations. As already advertised,
we will not consider left-right symmetric ones, which are purely
gravitational. Instead, we will switch the following $N=(1,0)$ world-sheet
supersymmetry compatible perturbation on
\begin{equation}
  \delta S_{\rm magnetic} = \frac{\sqrt{k k_G}H}{2\pi} \int {\rm
    d}^2 z \left(J^3 + \imath \psi^1 \psi^2\right) \bar J_G;
  \label{actmagdef}
\end{equation}
$\bar J_G$ being any Cartan current of the group factor $G$.

Although one may easily show the \emph{integrability} of this marginal
perturbation out of general arguments (see
\emph{e.g.}~\cite{Chaudhuri:1989qb}) it is instructive to pause and write an
explicit proof. If we limit ourselves to the bosonic sector, we can bosonize
the $\bar J_G$ current as $\bar J_G = \imath \bar \partial \varphi $ and interpret $\varphi
\left( z, \bar z\right)$ as an internal degree of freedom (see
App.~\ref{eom} for a more precise discussion). Incorporating the kinetic
term for the $\varphi$ field, the deformed action reads:
\begin{equation}
  S = S_{SU \left( 2\right)_k} \left( \alpha, \beta, \gamma \right) + \delta
  S_{\text{magnetic}} + \frac{k_G}{4 \pi } \int \di^2 z \, \partial \varphi \bar \partial
  \varphi,
\end{equation}
where $S_{SU \left( 2\right)_k}$ (Eq.~(\ref{SU2WZW})) now contains the
bosonic degrees of freedom only. The terms in previous expression can be
recollected so to give:
\begin{equation}
  S = S_{SU \left( 2\right)_k} \left( \alpha, \beta, \gamma +2 \sqrt{\frac{k_G}{k}} H
    \varphi \right) + \frac{k_G \left( 1 - 2H^2 \right)}{4 \pi }
  \int \di^2 z \, \partial \varphi \bar \partial \varphi,
\end{equation}
which is manifestly an exact \textsc{cft}. As a corollary, we observe that in the present
setting, $\mathcal{O}\left(\alpha^\prime\right)$ solutions of Eqs. (\ref{beta}) are
automatically promoted to all-order \emph{exact} solutions by simply
shifting $k \to k + 2$, just like for an ``ordinary'' \textsc{wzw} model.

The effect of the deformation at hand is to turn a (chromo)magnetic field on
along some Cartan direction inside $G$, which in turn induces a
gravitational back-reaction on the metric and the three-form antisymmetric
tensor. Following the previous discussion and App. \ref{eom} (\emph{i.e.} by
using Kaluza--Klein reduction), it is straightforward to read off the
space--time backgrounds from (\ref{SU2WZW}) and (\ref{actmagdef}). We
obtain:
\begin{equation}
  \di s^2 = \frac{k}{4}\left[\di \beta^2 + \sin^2
    \beta \di \alpha^2 + \left( 1 - 2 H^2\right)
    \left(\di \gamma + \cos \beta \di \alpha \right)^2
  \right]\label{S3metdef}
\end{equation}
and
\begin{equation}
  A = \sqrt{\frac{2k}{k_G}}
  H \left( \di \gamma  +  \cos \beta \di \alpha \right)
  \label{S3magdef}
\end{equation}
for the metric and gauge field, whereas neither the $B$-field nor the
dilaton are altered. The three-form field strength is however modified owing
to the presence of the gauge field (see Eqs.  (\ref{beta})):
\begin{equation}
  H_{[3]} = \di B - \frac{k_G}{4} A \land \di A = \frac{k}{4} \left(1 - 2
    H^2 \right) \sin \beta \di \alpha \land \di \beta \land \di
  \gamma
  \label{S3NSform}
\end{equation}
(the non-abelian structure of the gauge field plays no role since the
non-vanishing components are in the Cartan subalgebra\footnote{Similarly,
  the (chromo)magnetic field strength is given by $F=\di A$.}).

The deformed geometry (\ref{S3metdef}) is a \emph{squashed} three-sphere:
its volume decreases with respect to the original $S^3$ while its curvature
increases. These properties are captured in the expressions of the volume
form and Ricci scalar:
\begin{gather}
  \omega_{[3]}= \left( \frac{k}{4} \right)^{3/2}
  \sqrt{\left\vert 1 - 2 H^2\right\vert} \,
  \sin \beta \di \alpha \land \di \beta \land \di \gamma, \label{voldef} \\
  R = \frac{2}{k}(3 + 2 H^2).\label{ridef}
\end{gather}
The latter is constant and the background under consideration has $U(1) ×
SU(2)$ isometry generated by the Killing vectors $\set{L_3,R_1,R_2,R_3}$
whose explicit expression is reported in App.~\ref{sph},
Tab.~\ref{tab:su2-currents}.

This situation should be compared to the symmetric deformation generated by
the the marginal operator $\left(J^3 + i \psi^1 \psi^2\right) \bar J^3$. This is
purely gravitational and alters the metric, the $B$-field and the
dilaton~\cite{Giveon:1994ph}. The isometry is in that case broken to $U(1)×
U(1)$ and the curvature is not constant.

At this point one might wonder to what extent the constant curvature of the
asymmetric deformation is due to the large (almost maximal) residual
isometry $U(1) × SU(2)$. This question is answered in App.~\ref{def}, where
it is shown, in a general framework, that the isometry requirement is not
stringent enough to substantially reduce the moduli space of deformations.
In particular, the resulting curvature is in general not constant.

In the case under consideration, however, the geometric deformation is
driven by an integrable marginal perturbation of the sigma-model.  Combined
with the left-over isometry, this requirement leads to the above geometry
with constant curvature, Eq.~(\ref{ridef}).  From a purely geometrical point
of view (i.e. ignoring the \textsc{CFT} origin), such a deformation joins
the subclass of one-parameter families described in App.~\ref{def}, obtained
by demanding stability \emph{i.e.} integrability on top of the symmetry.

Notice finally that the $U(1) × SU(2)$ isometry originates from an affine
symmetry at the level of the sigma-model. The asymmetric marginal
deformation under consideration breaks the original affine $SU(2)_{\rm L}$
down to $U(1)_{\rm L}$, while it keeps the affine $SU(2)_{\rm R}$ unbroken.

It is worthwhile stressing that this asymmetric and the previously quoted
symmetric deformations of the three-sphere background are mutually
compatible and can be performed 
simultaneously~\cite{Kiritsis:1995ta,Kiritsis:1995iu}.

All the above discussion about integrability, geometry and isometries of the
$SU(2)_k$ magnetic perturbation is valid for the various asymmetric
deformations of $SL(2,\mathbb{R})$ that will be analysed in Sec.~\ref{ads}.


\boldmath
\subsection{Critical magnetic field and the geometric coset}\label{gens2}
\unboldmath

It was made clear in~\cite{Giveon:1994ph} that the symmetric deformation of
$SU(2)_k$ \textsc{wzw} is a well-defined theory for any value of the
deformation parameter. For an infinite deformation, the sigma-model becomes
a gauged \textsc{wzw} model $SU(2)_k /U(1)$ (\emph{bell} geometry) times a
decoupled boson. In some sense, the two $U(1)$ isometries present on the
deformation line act, for extreme deformation, on two disconnected spaces:
the bell and the real line.

As already stressed, the magnetic deformation of $SU(2)_k$ preserves a
larger symmetry, namely a $U(1) × SU(2)$, and has constant curvature. We are
in the framework discussed in App.~\ref{def}, Eqs. (\ref{defmet}) and
(\ref{defcon}) with $h = 2 H^2$. This deformation has an end-point where the
space is expected to factorize into a line with $U(1)$ isometry and a
two-dimensional constant-curvature space with $SU(2)$ isometry, which can
only be a two-sphere.

These statements can be made more precise by considering the background
(\ref{S3metdef}). The deformation parameter $H^2$ is clearly bounded: $H^2\leqslant
1/2$ (the boundary $H_{\text{max}}^2 = 1/2$ of the moduli space is
reminiscent of the $\mathrm{Im} \left(U\right) \to \infty$ limit in a
two-dimensional toroidal compactification).  In general, a three sphere can
be seen as an $S^1$-Hopf fibration over a base $S^2$. It is clear from
expressions (\ref{S3metdef}) and (\ref{voldef}) that the effect of the
magnetic field consists in changing the radius of the fiber. At $H^2 =
H_{\text{max}}^2$, this radius vanishes. 
The corresponding dimension decompactifies, and factorizes from the three-dimensional
geometry:
\begin{equation}
  S^3  \xrightarrow[H^2 \to H_{\text{max}}^2]{} \mathbb{R} ×
  S^2,
\end{equation}
where $S^2$ is the \emph{geometric coset} $SU(2)/ U(1)$. The $M_H
\to SU(2)/U(1)$ fibration trivializes in this limit. This can be
made more transparent by introducing a new coordinate:
\begin{equation}
  y = \sqrt{{k \over 2}\left({1\over 2} - H^2 \right)}\, \gamma.
\end{equation}
The metric and volume form now read:
\begin{equation}
  \di s^2 = \frac{k}{4}\left[\di \beta^2 + \sin^2
    \beta \di \alpha^2\right] + \di y^2
  + 2 \sqrt{{k \over 2}\left({1\over 2} - H^2 \right)} \, \cos
  \beta\,  \di \alpha  \di y  +
  {k \over 2}\left({1\over 2} - H^2 \right)\cos^2 \beta \di
  \alpha^2
 \label{S3metdefren}
\end{equation}
and
\begin{equation}
  \omega_{[3]}=\frac{k}{4}\, \sin \beta \di \alpha \land \di \beta
  \land \di y.
\end{equation}
For $H^2$ close to $H^2_{\text{max}}$, the $y$-direction factorizes
\begin{equation}
  \di s^2\xrightarrow[H^2 \to H^2_{\text{max}}]{}\di y^2 +
  \frac{k}{4}\left[\di \beta^2 + \sin^2 \beta \di
    \alpha^2\right],
\end{equation}
while the curvature ($R\to 8/k$) is entirely supported by the
remaining two-sphere of radius $\sqrt{k/4}$. The other background
fields read:
\begin{align}
  F &= \sqrt{\frac{2k}{k_G}} H \sin \beta \di \alpha \land \di \beta
  \xrightarrow[H^2 \to H^2_{\text{max}}]{} \sqrt{\frac{k}{k_G}} \sin \beta \di
  \alpha \land \di \beta,  \\
  H_{[3]} &= \sqrt{{k \over 2}\left({1\over 2} - H^2 \right)} \sin \beta \di \alpha
  \land \di \beta \land \di y \xrightarrow[H^2 \to H^2_{\text{max}}]{}
  0.
\end{align}

The above analysis deserves several comments. Our starting point
was a marginal deformation of the $SU(2)_k$ \textsc{wzw} model
embedded in heterotic strings and induced by a space--time
(chromo)magnetic field. Our observation is here that the
corresponding moduli space has a boundary, where the background is
$\mathbb{R} × S^2$, with finite magnetic field and no
three-form \textsc{NS} background. Being a marginal deformation,
this background is \emph{exact}, showing thereby that the
geometric coset is as good as a gauged \textsc{wzw} model
background. The latter appears similarly as the end-point of a
purely gravitational deformation; it carries neither magnetic
field nor $H_{[3]}$, but has a non-trivial dilaton.

Notice also that it was observed in the past that $S^2$ could
provide part of a string vacuum in the presence of \textsc{rr}
fluxes~\cite{Ferrara:1995ih}, but as usual when dealing with
\textsc{rr} fluxes, no exact conformal field theory description is
available.

The procedure we have developed so far for obtaining the two-sphere as an
exact background in the presence of a magnetic field is easily generalizable
to other geometric cosets of compact or non-compact groups. We will focus on
the latter case is Sec.~\ref{ads}, and analyze the electric/magnetic
deformations of AdS$_3$.

Our last comment concerns the quantization of the magnetic flux.
Indeed the two-sphere appears naturally  
as a factor of magnetic monopoles backgrounds~\cite{Kutasov:1998zh}.
At the limiting value of the deformation, the flux of the gauge
field through the two-sphere is given by:
\begin{equation}
  \mathcal{Q} = \int_{S^2} F = \sqrt{\frac{k}{k_G}} \int_{S^2}
  \hat\omega_2 = \sqrt{\frac{k}{k_G}} 4\pi,
  \label{flux}
\end{equation}
where $\hat \omega_2$ stands for the volume form of a unit-radius
two-sphere.  Therefore, a quantization of the magnetic charge
is only compatible with levels of the affine algebras such that:
\begin{equation}
  \label{chargeq}
  \frac{k}{k_G}=p^2 \ , \ \ p\in \mathbb{Z}.
\end{equation}
Actually, it was shown in~\cite{Johnson:1995kv} that the model
corresponding to the critical magnetic field can be obtained
directly with the following asymmetric gauged \textsc{wzw} model:
\begin{equation}
\frac{SU(2)_k ×  U(1)_{k_G}}{U(1)},
\end{equation}
where the left gauging lies in the $SU(2)_k$ \textsc{wzw} model
and the right gauging in the $U(1)_{k_G}$ of the gauge sector. The
cancellation of the anomalies of the coset dictates a condition on
the level of $SU(2)_k$ similar to~(\ref{chargeq}).


\subsection{Character formulas and modular invariance}
\label{pfs2}

We will here construct the contribution of the squashed three-sphere to the
partition function. This contribution is modular-covariant, and combines
with the remaining degrees of freedom into a modular-invariant result. Our
computation will also include the $S^2$ limiting geometry. We will consider
the case $k_G = 2$, \emph{i.e.} a $U(1)$ algebra generated by one
right-moving complex fermion.  We begin with the following combination of
$SU(2)_k$ supersymmetric characters and fermions from the gauge sector:\footnote{For 
convenience we choose here to denote by $k$ the level of the supersymmetric affine algebra; 
it contains a purely bosonic subalgebra at level $k-2$.}
\begin{equation}
  Z\oao{a ; h}{b; g} = \sum_{j,\bar{\jmath}=0}^{(k-2)/2} M^{j
    \bar{\jmath}} \ \chi^j \ \frac{\vartheta \oao{a}{b}}{\eta} \ \bar
  \chi^{\bar \jmath} \ \frac{\bar \vartheta \oao{h}{g}}{\bar \eta}.
\end{equation}
where the $\chi^j$'s are the characters of bosonic $SU(2)_{k-2}$, $(a,b)$ are
the $\zi_2$ boundary conditions for the left-moving fermions\footnote{We
  have removed the contribution of the fermion associated to $J^3$ since it
  is neutral in the deformation process.} and $(h,g)$ those of the
right-moving -- gauge-sector -- ones.  We can choose any matrix $M^{j \bar
  \jmath}$ compatible with modular invariance of $SU(2)_{k-2}$. Furthermore, the
supersymmetric $SU(2)_{k}$ characters can be decomposed in terms of those of
the $N=2$ minimal models:
\begin{equation}
  \chi^j (\tau ) \ \vartheta \oao{a}{b} (\tau , \nu ) = \sum_{m \in
    \zi_{2k}} \mc^{j}_{m} \oao{a}{b} \Theta_{m,k} \left( \tau,
    -\frac{2\nu}{k} \right),
\end{equation}
where the $N=2$ minimal-model characters, determined implicitly by this
decomposition, are given
in~\cite{Kiritsis:1988rv,Dobrev:1987hq,Matsuo:1987cj,Ravanini:1987yg}.

Our aim is to implement the magnetic deformation in this formalism. The
deformation acts as a boost on the left-lattice contribution of the Cartan
current of the supersymmetric $SU(2)_k$ and on the right current from the
gauge sector:
\begin{multline}
  \label{defspherespec}
  \Theta_{m,k} \ \bar \vartheta \oao{h}{g} = \sum_{n,\bar{n}} \mathrm{e}^{-\imath \pi
    g\left(\bar{n}+\frac{h}{2}\right)} q^{\frac{1}{2} \left(\sqrt{2k}
      n+\frac{m}{\sqrt{2k}}\right)^2}
  \bar{q}^{\frac{1}{2}\left(\bar{n}+\frac{h}{2}\right)^2}
  \\
  \longrightarrow \sum_{n,\bar{n}} \mathrm{e}^{-\imath\pi g\left(\bar{n}+\frac{h}{2}\right)}\ 
  q^{\frac{1}{2} \left[ \left(\sqrt{2k} n + \frac{m}{\sqrt{2k}} \right)\cosh
      x
      + \left( \bar n + \frac{h}{2} \right) \sinh x \right]^2} \\
  × \bar{q}^{\frac{1}{2} \left[ \left( \bar n + \frac{h}{2} \right) \cosh x
      +\left(\sqrt{2k} n + \frac{m}{\sqrt{2k}} \right)\sinh x \right]^2}.
\end{multline}
The boost parameter $x$ is related to the vacuum expectation value of the
gauge field as follows:
\begin{equation}
  \cosh x = \frac{1}{1-2H^2}. \label{defsphpar}
\end{equation}

We observe that, in the limit $H^2 \to H^2_{\mathrm{max}}$, the boost
parameter diverges ($x \to \infty$), and the following constraints arise:
\begin{equation}
  4(k+2)n+2m + 2 \sqrt{2k}\bar n + \sqrt{2k} h = 0.
  \label{chargcond}
\end{equation}
Therefore, the limit is well-defined only if the level of the supersymmetric
$SU(2)_k$ satisfies a quantization condition:
\begin{equation}
k = 2p^2 \ , \ \ p \in \zi.
\end{equation}
This is exactly the charge quantization condition for the flux of
the gauge field, Eq.~(\ref{chargeq}). Under this condition, the
constraints (\ref{chargcond}) lead to
\begin{subequations}
  \begin{align}
    m+ph &\equiv 0 \mod 2p =: 2pN,  \\
    \bar n &= 2  p n + N, \ \ N \in \zi_{2p}.
  \end{align}
\end{subequations}
As a consequence, the $U(1)$ corresponding to the combination of
charges orthogonal to~(\ref{chargcond}) decouples (its radius
vanishes), and can be removed. We end up with the following
expression for the $S^2$ partition function contribution:
\begin{equation}
  Z_{S^2}\oao{a ; h}{b ;
 g} = \sum_{j,\bar \jmath} M^{j \bar \jmath}
  \sum_{N \in \zi_{2p}} \mathrm{e}^{-\imath\pi g\left(N + \frac{h}{2}\right)}
  \ \mc^{j}_{p(2N-h)} \oao{a}{b} \ \bar{\chi}^{\bar
    \jmath}  ,\label{ZS2}
\end{equation}
in agreement with the result found in~\cite{Berglund:1996dv} by
using the coset construction. The remaining charge $N$ labels the
magnetic charge of the state under consideration. As a result, the
$R$-charges of the left $N=2$ superconformal algebra are:
\begin{equation}
\mathcal{Q}_{\mathrm{R}} = n + \frac{a}{2} - \frac{N-h/2}{p} \mod 2.
\end{equation}

We now turn to the issue of modular covariance. Under the
transformation $\tau \to - 1/\tau$, the minimal-model characters
transform as:
\begin{equation}
\mc^{j}_{m} \oao{a}{b} \left(-\frac{1}{\tau}\right) = \frac{1}{k} 
\mathrm{e}^{-\imath \frac{\pi}{2} ab} \sum_{j'
=0}^{(k-2)/2} \sin \left( \frac{\pi (2j+1)(2j'+1)}{k} \right)
\sum_{m' \in \zi_{2k}} \mathrm{e}^{\imath \pi \frac{mm'}{k}}
\mc^{j'}_{m'} \oao{b}{-a} (\tau).
\end{equation}
On the one hand, the part of the modular transformation related to $j$ is
precisely compensated by the transformation of
$\bar \chi^{\bar \jmath}$, in Eq. (\ref{ZS2}).  On the other hand, the part of the
transformation related to the spin structure $(a,b)$ is compensated by the
transformation of the other left-moving fermions in the full heterotic
string construction. We can therefore concentrate on the transformation
related to the $m$ charge, coming from the transformation of the
theta-functions at level $k$. We have
\begin{equation}
  \sum_{N \in \zi_{2p}} \mathrm{e}^{-\imath\pi g\left(N +
      \frac{h}{2}\right)} \ \mc^{j}_{p(2N-h)} \oao{a}{b} \to
  \frac{1}{\sqrt{2k}} \sum_{m' \in \zi_{4p^2}} \sum_{N \in \zi_{2p}}
  \mathrm{e}^{-\frac{\imath\pi}{2} \left(g+\frac{m'}{p}\right)h}
  \mathrm{e}^{2\imath\pi \frac{N(m'-pg)}{2p} } \mc^{j}_{m'}
  \oao{b}{-a};
\end{equation}
summing over $N$ in $\zi_{2p}$ leads to the constraint:
\begin{equation}
  m'- pg \equiv 0 \mod 2p := -2pN' \ , \ N' \in \zi_{2p}.
\end{equation}
So we end up with the sum
\begin{equation}
  \mathrm{e}^{\frac{\imath\pi}{2} hg} \sum_{N' \in \zi_{2p}}
  \mathrm{e}^{\imath\pi h\left(N'+\frac{g}{2}\right)}
  \mc^{j}_{p(2N'-g)} \oao{b}{-a}.
\end{equation}
combining this expression with the modular transformation of the remaining
right-moving fermions of the gauge sector, we obtain a modular invariant
result.

 In a similar way one can check the invariance of the full heterotic
string \textsc{cft} under $\tau \to \tau + 1$.



\section{Electric/magnetic deformations of \boldmath
$\mathrm{AdS}_3$ \unboldmath} \label{ads}

Anti-de-Sitter space in three dimensions is the (universal
covering of the) $SL(2,\mathbb{R})$ group manifold. It provides
therefore an exact string vacuum with \textsc{ns} background,
described in terms of the $SL(2,\mathbb{R})_k$ \textsc{wzw} model,
where time is embedded in the non-trivial geometry. We will
consider it as part of some heterotic string solution such as
$\mathrm{AdS}_3 \times S^3 \times T^4$ with \textsc{ns} three-form
field in $\mathrm{AdS}_3 \times S^3$ (near-horizon \textsc{ns}5/F1
background). The specific choice of a background is however of
limited importance for our purpose.

The issue of $\mathrm{AdS}_3$ deformations has been raised in several
circumstances. It is richer than the corresponding $S^3$ owing to the
presence of elliptic, hyperbolic or parabolic elements in
$SL(2,\mathbb{R})$. The corresponding generators are time-like, space-like
or light-like.  Similarly, the residual symmetry of a deformed
$\mathrm{AdS}_3$ has $U(1)$ factors, which act in time, space or light
direction.

Marginal \emph{symmetric} deformations of the
$SL(2,\mathbb{R})_k$ \textsc{wzw} are driven by bilinears $J\bar
J$ where both currents are in $SL(2,\mathbb{R})$ and are of the
same kind~\cite{Forste:1994wp,Israel:2003ry}. 
These break the $SL(2,\mathbb{R})_{\rm
L}\times SL(2,\mathbb{R})_{\rm R}$ affine symmetry to $U(1)_{\rm
L}\times U(1)_{\rm
  R}$ and allow to reach, at extreme values of the deformation, gauged
$SL(2,\mathbb{R})_k/U(1)$ \textsc{wzw} models with an extra free decoupled
boson. We can summarize the results as follows:
\renewcommand{\labelenumi}{(\alph{enumi})}
\begin{enumerate}
\item \underline{$J^3\bar J^3$} 
These are time-like currents (for conventions see
  App.~\ref{antids}) and the corresponding deformations connect
  $SL(2,\mathbb{R})_k$ with $U(1) \times SL(2,\mathbb{R})_k / U(1)\vert_{\rm
    axial\ or \ vector}$.  The $U(1)$ factor stands for a decoupled,
  non-compact time-like free boson\footnote{The extra bosons are always
    non-compact.}.  The gauged \textsc{wzw} model
  $SL(2,\mathbb{R})_k/U(1)\vert_{\rm axial}$ is the \emph{cigar}
  (two-dimensional Euclidean black hole) obtained by gauging the $g\to hgh$
  symmetry with the $h=\exp{i\frac{\lambda}{2}\sigma^2}$ subgroup, whereas
  $SL(2,\mathbb{R})_k/U(1)\vert_{\rm vector}$ corresponds to the $g\to
  hgh^{-1}$ gauging. This is the \emph{trumpet} and is T-dual to the
  cigar\footnote{Actually this statement holds only for the vector coset of
    the \emph{single cover} of $SL(2,\mathbb{R})$. Otherwise, from the n-th
    cover of the group manifold one obtains the n-th cover of the
    trumpet~\cite{Israel:2003ry}.}.  The generators of the affine residual
  symmetry $U(1)_{\rm L}\times U(1)_{\rm R}$ are both time-like (the
  corresponding Killing vectors are not orthogonal though). For extreme
  deformation, the time coordinate decouples and the antisymmetric tensor is
  trade for a dilaton. The isometries are time-translation invariance and
  rotation invariance in the cigar/trumpet.
\item \underline{$J^2\bar J^2$} 
The deformation is now induced by space-like currents.
  So is the residual affine symmetry $U(1)_{\rm L}\times U(1)_{\rm R}$ of
  the deformed model.  Extreme deformation points are T-dual: $U(1) \times
  SL(2,\mathbb{R})_k / U(1)$ where the $U(1)$ factor is space-like, and the
  $U(1)$ gauging of $SL(2,\mathbb{R})_k$ corresponds to $g\to hgh^{(-1)}$
  with $h=\exp{-\frac{\lambda}{2}\sigma^3}$~\cite{Dijkgraaf:1992ba}.
  The corresponding manifold is (some sector of) the
  Lorentzian two-dimensional black hole with a non-trivial dilaton.
  \item \underline{$(J^1 + J^3)(\bar J^1 + \bar J^3)$} 
  This is the last alternative, with
  both null currents. The deformation connects $\mathrm{AdS}_3$ with
  $\mathbb{R} \times
  \mathbb{R}^{1,1}$ plus a dilaton linear in the first factor. The $U(1)_{\rm
    L}\times U(1)_{\rm R}$ left-over current algebra is
    light-like\footnote{The isometry is actually richer by one (two
    translations plus a boost), but the extra generator (the boost) is not
    promoted to an affine symmetry of the sigma-model.}. Tensorized with 
    an $SU(2)_k$ CFT, this background describes a decoupling limit of the 
    \textsc{ns}5/F1 setup~\cite{Israel:2003ry}, where the fundamental strings 
    regularize the strong coupling regime.

\end{enumerate}

Our purpose here is to analyze \emph{asymmetric} deformations of
$\mathrm{AdS}_3$. Following App.~\ref{def} and the similar
analysis of Sec.~\ref{sec:squashing-sphere} for $S^3$, we expect
those deformations to preserve a $U(1)_{\rm L}\times
SL(2,\mathbb{R})_{\rm R}$ symmetry appearing as affine algebra from the
sigma-model point of view, and as isometry group for the
background. The residual $U(1)_{\rm L}$ factor can be time-like,
space-like or null depending on the current that has been used to
perturb the \textsc{wzw} model.

It is worth to stress that some deformations of $\mathrm{AdS}_3$ have been
studied in the past irrespectively of any conformal sigma-model or string
theory analysis. In particular it was observed in~\cite{Rooman:1998xf}, 
following~\cite{Reboucas:1983hn} that the 
three-dimensional\footnote{In fact, the original G\"odel solution is
four-dimensional, but the forth space dimension is a flat spectator. In
the following, we will systematically refer to the three-dimensional
non-trivial factor.}  G\"odel solution of Einstein equations could be
obtained as a member of a one-parameter family of $\mathrm{AdS}_3$
deformations that precisely enters the class we discuss in App.~\ref{def}.
G\"odel space-time 
is a constant-curvature Lorentzian manifold. Its isometry group
is $U(1) \times SL(2,\mathbb{R})$, and the $U(1)$ factor is generated by a
time-like Killing vector~;  these properties hold for generic values of the
deformation parameter. In fact the deformed $\mathrm{AdS}_3$ under
consideration can be embedded in a seven-dimensional flat space with
appropriate signature, as the intersection of four quadratic surfaces.
Closed time-like curves as well as high symmetry are inherited from the
multi-time maximally symmetric host space.  Another interesting property
resulting from this embedding is the possibility for changing the sign of
the curvature along the continuous line of deformation, without encountering
any singular behaviour (see Eq. (\ref{curnecogo})).

It seems natural to generalize the above results to \emph{new}
$\mathrm{AdS}_3$ deformations and promote them to \emph{exact} string
backgrounds. Our guideline will be the requirement of a $U(1) \times
SL(2,\mathbb{R})$ isometry group, with space-like or light-like $U(1)$'s,
following the procedure developed in App.~\ref{def}.

We will first review the time-like (elliptic) deformation of
$\mathrm{AdS}_3$ of~\cite{Rooman:1998xf} and recently
studied from a string perspective in~\cite{Israel:2003cx}. Hyperbolic
(space-like) and parabolic (light-like) deformations will be analyzed in
Secs. \ref{elec} and \ref{pp}.  All these deformations are of the type
(\ref{defmet}) and (\ref{defcon}) or (\ref{defmetnul}). We show in the
following how to implement these deformations as exact marginal
perturbations in the framework of the $SL(2,\mathbb{R})_k$ \textsc{wzw} model
embedded in heterotic string.

\boldmath
\subsection{Elliptic deformation: magnetic background}
\label{mag}
\unboldmath

Consider $\mathrm{AdS}_3$ in the $\left(\rho, t, \phi \right)$ coordinates, with
metric given in (\ref{eq:ads-rhotphi-metric}). In these coordinates, two
manifest Killing vectors are $L_3 \sim \d_t$ and $R_2 \sim \d_\phi$, time-like and
space-like respectively (see App.~\ref{antids},
Tab.~\ref{tab:currents-timelike}).

The deformation studied in~\cite{Rooman:1998xf} and quoted
as ``squashed anti de Sitter'' reads, in the above coordinates:
\begin{equation}
  \di s^2= \frac{L^2}{4} \left[ \di \rho^2 + \cosh^2 \rho  \di \phi^2 -
    \left( 1 + 2 H^2\right) \left( \di t + \sinh \rho \di
      \phi \right)^2 \right].
  \label{dsnecogo}
\end{equation}
It preserves a $U(1)\times SL(2,\mathbb{R})$ isometry group. The $U(1)$ is
generated by the \emph{time-like} vector $L_3$ of one original
$SL(2,\mathbb{R})$, while the right-moving $SL(2,\mathbb{R})$ is unbroken
(the expressions for the $\set{L_3,R_1,R_2,R_3}$ Killing vectors in
Tab.~\ref{tab:currents-timelike} remain valid at any value of the
deformation parameter). The Ricci scalar is constant
\begin{equation}
  R=-{2\over L^2}(3 - 2H^2), \label{curnecogo}
\end{equation}
while the volume form reads:
\begin{equation}
  \omega_{[3]} = \frac{L^3}{8} \sqrt{\left| 1+2H^2\right| }\, \cosh \rho\,  \di \rho
  \land \di \phi \land \di t.
\end{equation}
For $H^2 = 1/2$, this deformation coincides with the G\"odel
metric. It should be stressed, however, that nothing special
occurs at this value of the deformation parameter. The properties
of G\"odel space are generically reproduced at any $H^2>0$.

From a physical point of view, as it stands, this solution is pathological
because it has topologically trivial closed time-like curves through each
point of the manifold, like G\"odel space-time which belongs to this family.
Its interest mostly relies on the fact that it can be promoted to an exact
string solution, with appropriate \textsc{ns} and magnetic backgrounds. The
high symmetry of (\ref{dsnecogo}), is a severe constraint and, as was shown
in~\cite{Israel:2003cx}, the geometry at hand does indeed coincide with the
unique marginal deformation of the $SL(2,\mathbb{R})_k$ \textsc{wzw} that
preserves a $U(1)_{\rm L}\times SL(2,\mathbb{R})_{\rm R}$ affine algebra with
time-like $U(1)_{\rm L}$.

It is interesting to observe that, at this stage, the deformation parameter
$H^2$ \emph{needs not be positive.}:~(\ref{dsnecogo}) solves the
Einstein-Maxwell-scalar equations~\cite{Reboucas:1983hn} for any $H^2$.
Furthermore, for $H^2 <0$, there are no longer closed time-like
curves\footnote{As mentioned previously, the geometry at hand can be
  embedded in a seven-dimensional flat space, with signature $\varepsilon ---+++$,
  $\varepsilon = {\rm sign}(-H^2)$~\cite{Rooman:1998xf}.  This clarifies the origin
  of the symmetry as well as the presence or absence of closed time-like
  curves for positive or negative $H^2$.}. This statement is based on a
simple argument\footnote{This argument is local and must in fact be
  completed by global considerations on the manifold
  (see~\cite{Rooman:1998xf}).}.  Consider a time-like curve $x^\mu = x^\mu
\left( \lambda \right)$. By definition the tangent vector $\d_\lambda$ is
negative-norm, which, by using Eq.  (\ref{dsnecogo}), translates into
\begin{equation}
  \left( {\di\rho \over \di\lambda}\right)^2 + \cosh^2 \rho
  \left( {\di\phi \over \di\lambda}\right)^2 -
  \left( 1 + 2 H^2\right) \left( {\di t \over \di\lambda} + \sinh \rho
    {\di\phi \over \di\lambda} \right)^2 <0.
\end{equation}
If the curve is closed, $\di t /\di \lambda$ must vanish somewhere. At
the turning point, the resulting inequality,
\begin{equation}
  \left(2H^2 \sinh^2 \rho  -1\right) \left({\di\phi \over \di\rho}\right)^2
  >1
\end{equation}
is never satisfied for $H^2<0$, whereas it is for large enough\footnote{This
  means $\rho > \rho_{\rm c}$ where $\rho_{\rm c}$ is the radius where the norm of
  $\d_\phi$ vanishes and switches to negative ($\| \d_\phi \|^2 = L^2\left(
    1-2H^2\sinh^2 \rho \right)/4 $).  This never occurs for $H^2<0$.} $\rho$
otherwise.

This apparent \emph{regularization of the causal pathology},
unfortunately breaks down at the string level. In fact, as we will
shortly see, in order to be considered as a string solution, the
above background requires a (chromo)magnetic field. The latter
turns out to be proportional to $H$, and becomes
\emph{imaginary} in the range where the closed time-like curves
disappear. Hence, at the string level, unitarity is trade for
causality. It seems that no regime exists in the magnetic
deformation of $\mathrm{AdS}_3$, where these fundamental
requirements are simultaneously fulfilled.


In the heterotic backgrounds considered here, of the type
$\mathrm{AdS}_3 \times S^3 \times T^4$, the two-dimensional
$N=(1,0)$ world-sheet action corresponding to the $\mathrm{AdS}_3$
factor is
 \begin{equation}
    S_{SL(2,\mathbb{R})_k} =\frac{1}{2\pi} \int {\rm d}^2 z
    \left\{{k\over 4}
      \left( \d \rho \db \rho - \d t \db t + \d \phi \db \phi - 2 \sinh \rho \,
        \d \phi \db t \right) + \eta_{ab}\, \psi^a \db \psi^b\right\},
    \label{SL2RWZW}
  \end{equation}
where $\eta_{ab}={\rm diag \ }(++-)$, $a=1,2,3$ and $\psi^a$ are
the left-moving superpartners of the $SL(2,\mathbb{R})_k$ currents
(see Tab.~\ref{tab:currents-timelike}). The corresponding
background fields are the metric (Eq.
(\ref{eq:ads-rhotphi-metric})) with radius $L=\sqrt{k}$ and the NS
B-field:
\begin{equation}
 \label{eq:ads-rhotphi-B}
 B = - \frac{k}{4} \sinh \rho \di \phi \land \di t.
\end{equation}
The three-form field strength is $H_{[3]} = \di
B=-\frac{2}{\sqrt{k}}\,
 \omega_{[3]}$ with $\omega_{[3]}$ displayed in Eq.~\eqref{eq:ads-rhotphi-vf}.

The asymmetric perturbation that preserves a $U(1)_{\rm L}\times
SL(2,\mathbb{R})_{\rm  R}$ affine algebra with time-like
$U(1)_{\rm L}$ is $\delta S_{\rm magnetic}$ given in Eq.
(\ref{actmagdef}), where $J^3$ now stands for the left-moving
time-like $SL(2,\mathbb{R})_k$ current given in App.~\ref{antids},
Tab.~\ref{tab:currents-timelike}. This perturbation corresponds to
switching on a (chromo)magnetic field, like in the $SU(2)_k$
studied in Sec.~\ref{sphere}. It is marginal and can be integrated
for finite values of $H$, and is compatible with the $N=(1,0)$
world-sheet supersymmetry. The resulting background fields,
extracted in the usual manner from the deformed action
 are the metric (\ref{dsnecogo}) with
radius $L=\sqrt{k}$ and the following gauge field:
\begin{equation}
  A = H \sqrt{\frac{2k}{k_g}} \left(\di t + \sinh \rho \di \phi
  \right).
  \label{adsmag}
\end{equation}
The NS $B$-field is not altered by the deformation, (Eq.
(\ref{eq:ads-rhotphi-B})), whereas the three-form field strength
depends explicitly on the deformation parameter $H$, because of
the gauge-field contribution:
\begin{equation}
  H_{[3]} = \di B - \frac{k_G}{4} A \land \di A =
 - \frac{k}{4}\left( 1+ 2H^2\right) \cosh \rho \di \rho \land \di \phi \land \di
 t.
  \label{adsmagH}
\end{equation}

One can easily check that the background fields (\ref{dsnecogo}),
(\ref{adsmag}) and (\ref{adsmagH}) solve the lowest-order
equations of motion (\ref{beta}). Of course the solution we have
obtained is exact, since it has been obtained as the marginal
deformation of an exact conformal sigma-model. The interpretation
of the deformed model in terms of background fields $\{ G_{ab},
B_{ab}, F_{ab}^G \}$ receives however the usual higher-order
correction summarized by the shift $k \to k + 2 $ as we have
already seen for the sphere in Sec.~\ref{sec:squashing-sphere}.

Let us finally mention that it is possible to extract the spectrum
and write down the partition function of the above
theory~\cite{Israel:2003cx}, since the latter is an exact
deformation of the $SL(2,\mathbb{R})_k$ \textsc{wzw} model. This
is achieved by deforming the associated elliptic Cartan
subalgebra. The following picture emerges then from the analysis
of the spectrum. The short-string spectrum, corresponding to
world-sheets trapped in the center of the space--time (for some
particular choice of coordinates) is well-behaved, because these
world-sheets do not feel the closed time-like curves which are
``topologically large''. On the contrary, the long strings can
wrap the closed time-like curves, and their spectrum contains many
tachyons. Hence, the caveats of G\"odel space survive the string
framework, at any value of $H^2>0$. One can circumvent them by
slightly deviating from the G\"odel line with an extra purely
gravitational deformation, driven by $J^3 \bar{J}^3$. This
deformation isolates the causally unsafe region, $\rho > \rho_{\rm
c}$ (see~\cite{Israel:2003cx} for details). It is similar in 
spirit with the supertubes domain-walls of~\cite{Drukker:2003sc} 
curing the G\"odel-like space-times with RR backgrounds.

As already stressed,
one could alternatively switch to negative $H^2$. Both metric and
antisymmetric tensor are well-defined and don't suffer of causality
problems. The string picture however
breaks down because the magnetic field (Eq. (\ref{adsmag}))
becomes imaginary.

\boldmath
\subsection{Hyperbolic deformation: electric background}
\label{elec} \unboldmath

\subsubsection{The background and its CFT realization}

We will now focus on a different deformation. We use coordinates
(\ref{eq:ads-rxt-coo}) with metric (\ref{eq:ads-rxt-met}), where the
manifest Killing vectors are $L_2 \sim \d_x$ (space-like) and $R_3 \sim \d_\tau$
(time-like) (see App.~\ref{antids}, Tab.~\ref{tab:currents-spacelike}). This
time we perform a deformation that preserves a $SO(1,1) \times SL(2,\mathbb{R})$
isometry. The $SO(1,1)$ corresponds to the non-compact space-like Killing vector $L_2$,
whereas the $SL(2,\mathbb{R})$ is generated by $R_1, R_2, R_3$, which are
again not altered by the deformation. This is achieved by implementing Eqs.
(\ref{defmet}) and (\ref{defcon}) in the present set up, with $\xi = \d_x$
and $h = 2H^2$. The resulting metric reads:
\begin{equation}
  \di s^2= \frac{L^2}{4}\left[ \di r^2 - \cosh^2 r \di \tau^2 +
    \left( 1-2H^2\right) \left( \di x + \sinh r \di \tau \right)^2\right].
  \label{dsnecoma}
\end{equation}
The scalar curvature of this manifold is constant
\begin{equation}
  R=-\frac{2}{L^2}\left(3+2H^2\right)\label{Rel}
\end{equation}
and the volume form
\begin{equation}
  \omega_{[3]} = \frac{L^3}{8}\sqrt{\left|1-2H^2 \right|}\, \cosh^2 r\,  \di r
  \land \di \tau \land \di x.\label{vfelec}
\end{equation}

Following the argument of Sec.~\ref{mag}, one can check whether
closed time-like curves appear. Indeed, assuming their existence,
the following inequality must hold at the turning point
\emph{i.e}. where $\di t/\di \lambda$ vanishes ($\lambda$ being
the parameter that describes the curve):
\begin{equation}
  \left( 2H^2 -1 \right)\left( {\di x \over \di r} \right)^2>1 .
\end{equation}
The latter cannot be satisfied in the regime $H^2<1/2$. Notice
that the manifold at hand is well behaved, even for negative
$H^2$.

Let us now leave aside these questions about the classical
geometry, and address the issue of string realization of the above
background. As already advertised, this is achieved by considering
a world-sheet-supersymmetric marginal deformation of the
$SL(2,\mathbb{R})_k$ \textsc{wzw} model that implements
(chromo)electric field. Such a deformation is possible in the
heterotic string at hand:
\begin{equation}
  \delta S_{\rm electric} = \frac{\sqrt{k k_G}H}{2\pi} \int {\rm
    d}^2 z \left(J^2 + i \psi^1 \psi^3\right) \bar J_G,
\label{actelecdef}
\end{equation}
($\bar J_G$ is any Cartan current of the group $G$ and $J^2$ is
given in App.~\ref{antids}, Tab.~\ref{tab:currents-spacelike}),
and corresponds, as in previous cases, to an integrable marginal
deformation. The deformed conformal sigma-model can be analyzed in
terms of background fields. The metric turns out to be
(\ref{dsnecoma}), whereas the gauge field and three-form tensor
are
\begin{align}
  A &= H \sqrt{\frac{2k}{k_g}} \left( \di x + \sinh r \di \tau
  \right),  \\
  H_{[3]} &= \frac{k}{4} \left(1-2H^2\right) \cosh r \di r \land \di \tau \land \di
  x.
\end{align}
As expected, these fields solve Eqs. (\ref{beta}).

The background under consideration is a new string solution generated as a
hyperbolic deformation of the $SL(2,\mathbb{R})_k$ \textsc{wzw} model. In
contrast to what happens for the elliptic deformation (magnetic background
analyzed in Sec.~\ref{mag}), the present solution is perfectly sensible,
both at the classical and at the string level. 

\subsubsection{The spectrum of primaries}

The electric deformation of $\mathrm{AdS}_3$ is an exact string background.
The corresponding conformal field theory is however more difficult to deal
with than the one for the elliptic deformation. In order to write down its
partition function, we must decompose the $SL(2,\mathbb{R})_k$ partition
function in a hyperbolic basis of characters, where the implementation of
the deformation is well-defined and straightforward; this is a notoriously
difficult exercise. On the other hand the spectrum of primaries is
known\footnote{In the following we do not consider the issue of the
  spectral-flow representations. The spectral-flow symmetry is apparently
  broken by the deformation considered here.}  from the study of the
representations of the Lie algebra in this basis (see
\emph{e.g.}~\cite{Vilenkin}, and~\cite{Dijkgraaf:1992ba} for the spectrum of 
the hyperbolic gauged \textsc{wzw} model). 
The part of the heterotic spectrum of interest
contains the expression for the primaries of $N=(1,0)$ affine
$SL(2,\mathbb{R})$ at purely bosonic level\footnote{More precisely we
  consider primaries of the purely bosonic affine algebra with an arbitrary
  state in the fermionic sector.} $k+2$, together with some $U(1)$ from the
lattice of the heterotic gauge group:
\begin{align}
  L_0 &= -\frac{j(j-1)}{k} - \frac{1}{2} \left(n+\frac{a}{2}\right)^2,
  \label{leftspecads} \\
  \bar L_0 &= -\frac{j(j-1)}{k} +   \frac{1}{2}
  \left(\bar{n}+\frac{h}{2}\right)^2  ,
  \label{rightspecads}
\end{align}
where the second Casimir of the representation of the $SL(2,\mathbb{R})$
algebra, $-j(j-1)$, explicitly appears. The spectrum contains
\emph{continuous representations}, with $j = \frac{1}{2} + \imath s$, $s \in
\mathbb{R}_+$. It also contains \emph{discrete representations}, with $j \in
\mathbb{R}_{+}$, lying within the unitarity range $1/2 < j < (k+1)/2$
(see~\cite{Maldacena:2000hw, Petropoulos:1990fc}). In both cases the
spectrum of the hyperbolic generator $J^2$ is $\mu \in \mathbb{R}$. The
expression for the left conformal dimensions, Eq.~(\ref{leftspecads}), also
contains the contribution from the world-sheet fermions associated to the
$\imath \psi^1 \psi^3$ current. The sector (\textsc{r} or \textsc{ns}) is labelled
by $a \in \zi_2$. Note that the unusual sign in front of the lattice is the
natural one for the fermions of the light-cone directions. In the
expression~(\ref{rightspecads}) we have similarly the contribution of the
fermions of the gauge group, where $h$ labels the corresponding sector.

We are now in position to follow the procedure, familiar from the previous
examples: we have to (\emph{i}) isolate from the left spectrum the lattice
of the supersymmetric hyperbolic current $J^2 + \imath \psi^1 \psi^3$ and
(\emph{ii}) perform a boost between this lattice and the fermionic lattice
of the gauge field. We hence obtain the following expressions:
\begin{align}
  \begin{split}
    L_0 &= -\frac{j(j-1)}{k} - \frac{\mu^2}{k+2} - \frac{k+2}{2k}
    \left( n+ \frac{a}{2} + \frac{2\mu}{k+2} \right)^2 +\\
    & \hspace{6em}+ \frac{1}{2} \left[ \sqrt{\frac{2}{k}}
      \left(\mu + n + \frac{a}{2} \right) \cosh x  + \left(\bar{n} +
        \frac{h}{2} \right) \sinh x \right]^2,
  \end{split}\label{leftspecadsdef}\\
  \bar L_0 &= -\frac{j(j-1)}{k} + \frac{1}{2} \left[ 
    \left(\bar{n} + \frac{h}{2} \right) \cosh x  +\sqrt{\frac{2}{k}}
    \left(\mu + n + \frac{a}{2} \right)  \sinh x  \right]^2.
  \label{rightspecadsdef}
\end{align}
The relation between the boost parameter $x$ and the deformation parameter
$H^2$ is given in Eq. (\ref{defsphpar}), as for the case of the $SU(2)_k$
deformation. In particular it is worth to remark that the first three terms
of~(\ref{leftspecadsdef}) correspond to the left weights of the
worldsheet-supersymmetric two-dimensional Lorentzian black hole, \emph{i.e.} the
$SL(2,\mathbb{R})/ O(1,1)$ gauged super-\textsc{wzw} model.

\boldmath
\subsection{Parabolic deformation: electromagnetic-wave background}
\label{pp}
\unboldmath

In the deformations of Secs. \ref{mag} and \ref{elec}, one
$SL(2,\mathbb{R})$ isometry breaks down to a $U(1)$ generated either by a
time-like or by a space-like Killing vector.  Deformations which preserve a
light-like isometry do also exist and are easily implemented in Poincar\'{e}
coordinates.

We require that the isometry group is $SO(1,1) \times SL(2,\mathbb{R})$ with a null
Killing vector for the $SO(1,1)$ factor. Following the deformation procedure
described in App.~\ref{def} for the particular case of light-like residual
isometry, Eq. (\ref{defmetnul}) with $h=2H^2$, we are lead to
\begin{equation}
  \di s^2 =L^2\left[\frac{ \di u^2 }{u^2}+ \frac{\di x^+ \di x^-
    }{u^2}-2H^2 \left(\frac{\di x^+}{u^2}\right)^2 \right].
  \label{dsemdef}
\end{equation}
The light-like $SO(1,1)$ Killing vector is $L_1 + L_3 \sim \d_-$ (see
App.~\ref{antids},  Tab.~\ref{tab:currents-poincare}). The
remaining $SL(2,\mathbb{R})$ generators are
$\set{R_1+R_3,R_1-R_3,R_2}$ and remain unaltered after the
deformation.

The above deformed anti-de-Sitter geometry looks like a
superposition of $\mathrm{AdS}_3$ and of a plane wave. As usual,
the sign of $H^2$ is free at this stage, and $H^2<0$ are equally
good geometries. In the near-horizon region ($\vert u \vert \gg
\left\vert H^2\right\vert$) the geometry is not sensitive to the
presence of the wave. On the contrary, this plane wave dominates
in the opposite limit, near the conformal boundary. 

The volume form is not affected by the deformation, and it is
still given in (\ref{eq:ads-poinc-volume}); neither is the Ricci
scalar modified:
\begin{equation}
  R=-\frac{6}{L^2} . \label{Remdef}
\end{equation}
Notice also that the actual value of $\vert H \vert$ is not of physical
significance: it can always be absorbed into a reparameterization $x^+ \to
x^+ /\vert H \vert$ and $x^-\to x^- \vert H \vert$. The only relevant values
for $H^2$ can therefore be chosen to be $0, \pm 1$.

We now come to the implementation of the geometry (\ref{dsemdef})
in a string background. The only option is to perform an
asymmetric exactly marginal deformation of the heterotic
$SL(2,\mathbb{R})_k$ \textsc{wzw} model that preserves a
$U(1)_{\rm L}\times SL(2,\mathbb{R})_{\rm R}$ affine symmetry.
This is achieved by introducing
\begin{equation}
  \delta S_{\rm electric-magnetic} = -4{\sqrt{k k_G}H} \int
  \di^2 z \left(J^1 + J^3 + i \left( \psi^1 + \psi^3 \right)
    \psi^2 \right) \bar J_G,
\label{actemdef}
\end{equation}
($J^1 + J^3$ is defined in App.~\ref{antids},
Tab.~\ref{tab:currents-poincare}). The latter perturbation is
integrable and accounts for the creation of an
(chromo)electromagnetic field 
\begin{equation}
  A = 2 \sqrt{2k\over k_G} H {\di x^+\over u^2}.\label{adsem}
\end{equation}
It generates precisely the deformation (\ref{dsemdef}) and leaves
unperturbed the \textsc{ns} field, $H_{[3]}= \di B = -
\frac{2}{\sqrt{k}} \, \omega_{[3]}$.

As a conclusion, the $\mathrm{AdS}_3$ plus plane-wave
gravitational background is described in terms of an exact
conformal sigma model, that carries two extra background fields: a
\textsc{ns} three-form and an electromagnetic two-form. Similarly
to the symmetric parabolic deformation~\cite{Israel:2003ry}, the
present asymmetric one can be used to construct a space--time
supersymmetric background. The $SL(2,\mathbb{R})_k$-\textsc{cft}
treatment of the latter deformation would need the knowledge of
the parabolic characters of the affine algebra, not available at
present.

As already stressed for the elliptic deformation (end of Sec.
\ref{mag}), the residual affine symmetry leaves the possibility
for an extra, purely gravitational, symmetric marginal
deformation. Although the systematic analysis of the full
$\mathrm{AdS}_3$ landscape is beyond the present scope, we would
like to quote the effect of such a deformation on the parabolic
line. The perturbation which is turned on is $\sim \int {\rm d}^2
z \ J^+ \bar J^+$ (the currents are given in Eqs.
(\ref{eq:currents-SL2R}) and Tab. \ref{tab:currents-poincare}),
with parameter $1/M^2$. In the absence of electromagnetic
background ~\cite{Israel:2003ry}, this deformation connects the
\textsc{ns}5/F1 background to the pure \textsc{ns}5 dilatonic solution. Here it is
performed on top of the asymmetric one, which introduces an
electromagnetic wave, and we find:
\begin{subequations}
  \begin{align}
     \di s^2 &=k\left[\frac{ \di u^2 }{u^2}+ \frac{\di x^+ \di x^-
    }{u^2+1/M^2}-2H^2 \left(\frac{\di x^+}{u^2+1/M^2}\right)^2 \right],\\
    B &= \frac{k}{2} \frac{1}{u^2 + 1/M^2} \di x^+ \land \di x^-,\\
   \rm{e}^{2\Phi} &= k g_{\rm s}^2 \frac{u}{u^2 +1/M^2},\\
\intertext{plus an electromagnetic field:}
   A &= 2 \sqrt{2k\over k_G} H {\di x^+\over u^2 + 1/M^2}.
  \end{align}
\end{subequations}
At $M^2 \to \infty$, we recover the solution (\ref{dsemdef}) and
(\ref{adsem}), whereas at $M^2 \to 0$ the present solution
asymptotes the linear dilaton background. Therefore, in an \textsc{ns}5/F1
setup, the deformation at hand may be relevant for investigating
the holography of little string theories~\cite{Aharony:1998ub}.

\subsection{A remark on discrete identifications}

Before closing the chapter on $\mathrm{AdS}_3$, we would like to
discuss briefly the issue of discrete identifications. So far we
have focused on continuous deformations as a procedure for
generating new backgrounds. It appeared that under specific
symmetry and integrability requirements, the moduli of such
deformations are unique, and the corresponding backgrounds are
described in terms of exact two-dimensional conformal models.

In the presence of isometries, discrete identifications provide
alternatives for creating new backgrounds. Those have the
\emph{same} local geometry --~except at possible fixed points~--
but differ with respect to their
global properties. Whether these identifications can be
implemented as orbifolds, at the level of the underlying
two-dimensional model is very much dependent on each specific
case.

For $\mathrm{AdS}_3$, the most celebrated geometry obtained by discrete
identification is certainly the \textsc{btz} black
hole~\cite{Banados:1992wn}. The discrete identifications are made along the
integral lines of the following Killing vectors (defined in Eqs.
(\ref{eq:Killing-SL2R})):
\begin{subequations}
  \begin{align}
     \text{extremal case}&: \ \ \xi = 2\imath r_+  R_2 -
     \imath\left( R_1-R_3\right) -
     \imath \left( L_1+L_3\right) \label{ext},\\
    \text{non-extremal case}&: \ \ \xi' =
    \imath \left(r_+ + r_- \right)R_2 - \imath \left(r_+ - r_- \right)L_2,\label{next}
  \end{align}
\end{subequations}
where $r_+$ and $r_-$ are the outer and inner horizons, coinciding
for the extremal black hole. Many subtleties arise, which concern
\emph{e.g.} the appearance of closed time-like curves; a
comprehensive analysis of these issues can be found in~\cite{Banados:1993gq}.
At the string theory level, this projection is realized as an \emph{orbifold}, 
which amounts to realize the projection of the string spectrum onto invariant
states and to add twisted sectors~\cite{Natsuume:1998ij,Hemming:2001we}.

Besides the \textsc{btz} solution, other locally $\mathrm{AdS}_3$ geometries
are obtained, by imposing identification under purely left (or right)
isometries, refereed to as self-dual (or anti-self-dual) metrics. These were
studied in~\cite{Coussaert:1994tu}. Their classification and isometries are
exactly those of the asymmetric deformations studied in the present chapter.
The Killing vector used for the identification is (A) time-like (elliptic),
(B) space-like (hyperbolic) or (C) null (parabolic), and the isometry group
is $U(1) \times SL(2,\mathbb{R})$. It was pointed out in~\cite{Coussaert:1994tu}
that the resulting geometry was free of closed time-like curves only in the
case (B).

We could clearly combine the continuous deformations with the
discrete identifications -- whenever these are compatible -- and
generate thereby new backgrounds. This offers a large variety of
possibilities that deserve further investigation (issue of
horizons, closed time-like curves \dots). One can \emph{e.g.}
implement the non-extremal \textsc{btz} identifications (\ref{next}) on the
hyperbolic continuous deformation (\ref{dsnecoma}) since the
isometry group of the latter contains the vectors of the former.

Furthermore, it can be used to generate new interesting solutions of
Einstein equations by performing discrete identifications in the spirit
of~\cite{Coussaert:1994tu}. In the latter, the residual isometry group was
precisely the one under consideration here, so that our deformation is
compatible with their discrete identification.

Similarly, the extremal \textsc{btz} identifications (\ref{ext}) are
compatible with the isometries of the parabolic deformation
(\ref{dsemdef}). One could thus create extremal black holes out of
this AdS/plane-wave solution.

\section{Limiting geometries: \boldmath $\mathrm{AdS}_2$ and $H_2$
 \unboldmath}
\label{gencos}

\boldmath
\unboldmath

We have analyzed in Sec.~\ref{gens2} the behaviour of the magnetic
deformation of $SU(2)_k$, at some critical (or boundary) value of
the modulus $H^2$, where the background factorizes as
$\mathbb{R}\times S^2$ with vanishing \textsc{ns} three-form and
finite magnetic field. We would like to address this question for
the asymmetric deformations of the $SL(2,\mathbb{R})_k$ model and
show the existence of limiting situations where the geometry
indeed factorizes, in agreement with the expectations following
the general analysis of App.~\ref{def}.

In general, exact deformations of string backgrounds as those we
are considering here, are carried by a modulus that controls the
string spectrum. The modulus might exhibit critical or boundary
values, where a whole sector of states becomes massless or
infinitely massive, and decouples. Such a phenomenon corresponds
to the decompactification of some compact coordinate, which
decouples from the remaining geometry. This is exactly what
happens for the magnetic deformation of $SU(2)_k$ \textsc{wzw}
where the $S^3$ is more and more squashed, and eventually shrinks
to a $\mathbb{R}\times S^2$. Not only is the geometry affected,
but the antisymmetric tensor disappears in this process, and the
$S^2$ is left with a finite magnetic field that ensures the
consistency of the string theory.

What can we expect in the framework of the $SL(2,\mathbb{R})_k$
asymmetric deformations? Any limiting geometry must have the
generic $U(1) \times SL(2,\mathbb{R})_k$ isometry that translates
the affine symmetry of the conformal model. If a line decouples,
it accounts for the $U(1)$, and the remaining two-dimensional
surface must be $SL(2,\mathbb{R})$-invariant.  Three different
situations may arise: $\mathrm{AdS}_2$, $H_2$ or dS$_2$.  Anti de
Sitter in two dimensions is Lorentzian with negative curvature;
the hyperbolic plane $H_2$ (also called Euclidean anti de Sitter)
is Euclidean with negative curvature; de Sitter space is
Lorentzian with positive curvature. 

Three deformations are available for $\mathrm{AdS}_3$ and these
have been analyzed in Sec.~\ref{ads}. For unitary string theory,
all background fields must be real and consequently $H^2>0$ is the
only physical regime. In this regime, only the hyperbolic
(electric) deformation exhibits a critical behaviour at $H^2_{\rm
max} = 1/2$. For $H^2 < 1/2$, the deformation at hand is a
Lorentzian manifold with no closed time-like curves (see
Sec.~\ref{elec}).  When $H^2 > 1/2$, $\det {\bf g}>0$ and
\emph{two} time-like directions appear. At $H^2_{\vphantom m} =
H^2_{\rm max}$, $\det {\bf g}$ vanishes, and this is the signature
that some direction indeed decompactifies.

We proceed therefore as in Sec.~\ref{gens2}, and define a rescaled
coordinate in order to keep the decompactifying direction into the
geometry and follow its decoupling:
\begin{equation}
  y=\sqrt{\frac{k}{2}\left(\frac{1}{2}  -H^2\right)} \, x\ .
\end{equation}
The metric and volume form now read:
\begin{equation}
  \di s^2 = \di y^2 + \frac{k}{4} \left[ \di r^2 - \left( 1 + 2 H^2 \sinh^2 r \right)
    \di \tau^2 \right] + \sqrt{k\left(1-2H^2 \right)}\sinh r\, \di \tau\, \di y
  \label{ads3metdefren}
\end{equation}
and
\begin{equation}
  \omega_{[3]} = \frac{k}{4} \, \cosh r\,  \di r
  \land \di \tau \land \di y.
\end{equation}
For $H^2_{\vphantom x}$ close to $H^2_{\rm max}$, the $y$-direction
factorizes
\begin{equation}
  \di s^2\xrightarrow[H^2 \to H^2_{\text{max}}]{}\di y ^2 +  \frac{k}{4}\left[\di
    r^2- \cosh^2 r \, \di \tau^2\right].
\end{equation}
The latter expression captures the phenomenon we were expecting:
\begin{equation}
  {\rm AdS}_3\xrightarrow[H^2 \to H^2_{\text{max}}]{} \mathbb{R} \times
  {\rm AdS}_2.
\end{equation}
It also shows that the two-dimensional anti de Sitter has radius $\sqrt{k/4}$
and supports entirely the curvature of the limiting geometry, $R=-8/k$ (see
expression (\ref{Rel})).

The above analysis shows that, starting from the
$SL(2,\mathbb{R})_k$ \textsc{wzw} model, there is a line of
continuous exact deformation (driven by a (chromo)electric field)
that leads to a conformal model at the boundary of the modulus
$H^2$. This model consists of a free non-compact boson times a
geometric coset $\mathrm{AdS}_2\equiv SL(2,\mathbb{R})/U(1)$, with
a finite electric field:
\begin{equation}
  F=  \sqrt{k \over k_G}\cosh r \, \di r \land \di \tau
\end{equation}
and vanishing \textsc{ns} three-form background. The underlying
geometric structure that makes this phenomenon possible is that
$\mathrm{AdS}_3$ can be considered as a non-trivial $S^1$
fibration over an $\mathrm{AdS}_2$ base. The radius of the fiber
couples to the electric field, and vanishes at $H^2_{\rm max}$.
The important result is that this enables us to promote the
geometric coset $\mathrm{AdS}_2$ to an exact string vacuum.

We would like finally to comment on the fate of dS$_2$ and $H_2$
geometries, which are both $SL(2,\mathbb{R})$-symmetric. De Sitter
and hyperbolic geometries are not expected to appear in physical
regimes of string theory. The $H_3$ sigma-model, for example, is
an exact conformal field theory, with imaginary antisymmetric
tensor background though~\cite{Gawedzki:1991yu,Teschner:1997ft}. 
Similarly, imaginary \textsc{ns}
background is also required for de Sitter vacua to solve the
low-energy equations (\ref{beta}). It makes sense therefore to
investigate regimes with $H^2<0$, where the electric or magnetic
backgrounds are indeed imaginary.

The elliptic (magnetic) deformation studied in Sec.~\ref{mag} exhibits a
critical behaviour in the region of negative $H^2$, where the geometry does
not contain closed time-like curves. The critical behaviour appears at the
minimum value $H^2_{\rm min} =-1/2$, below which the metric becomes
Euclidean. The vanishing of $\det{\bf g}$ at this point of the deformation
line, signals the decoupling of the time direction. The remaining geometry
is nothing but a two-dimensional hyperbolic plane $H_2$. It is Euclidean
with negative curvature $R=-8/k$ (see Eq.  (\ref{curnecogo}) with $L^2 =
k$).

All this can be made more precise by introducing  a rescaled time
coordinate:
\begin{equation}
  T =\sqrt{ \frac{k}{2} \left( \frac{1}{2}  +H^2\right)} \, t.
\end{equation}
The metric and volume form now read:
\begin{equation}
  \di s^2= - \di T^2 + \frac{k}{4} \left[ \di \rho^2 + \left( 1 -
      2 H^2 \sinh^2 \rho \right) \di \phi^2 \right] -
  \sqrt{k\left(1+2H^2\right)} \sinh \rho \di \phi \di T
  \label{ads3metmagdefren}
\end{equation}
and
\begin{equation}
  \omega_{[3]} =  \frac{k}{4}  \cosh \rho  \di \rho \land \di \phi  \land \di T.
\end{equation}
For $H^2_{\vphantom x}$ close to $H^2_{\rm min}$, the
$T$-direction factorizes
\begin{equation}
  \di s^2\xrightarrow[H^2_{\vphantom m} \to H^2_{\text{max}}]{} -\di
  T ^2 +  \frac{k}{4}\left[ \di \rho^2 + \cosh^2 \rho \di \phi^2 \right].
\end{equation}
The latter expression proves the above statement:
\begin{equation}
  {\rm AdS}_3\xrightarrow[H^2_{\vphantom m} \to H^2_{\text{min}}]{}
  \mathbb{R} \times H_2,
\end{equation}
and the two-dimensional hyperbolic plane has radius $\sqrt{k/4}$.

Our analysis finally shows that the continuous line of exactly marginal
(chromo)magnetic deformation of the $SL(2,\mathbb{R})$ conformal model,
studied in Sec.~\ref{mag}, has a boundary at $H^2 = -1/2 $ where its target
space is a free time-like coordinate times a hyperbolic plane. The price to
pay for crossing $H^2 = 0$ is an imaginary magnetic field, which at $H^2 =
-1/2$ reads:
\begin{equation}
  F= \sqrt{-{k \over k_G}}\cosh \rho \, \di \phi \land \di \rho.
\end{equation}
The \textsc{ns} field strength vanishes at this point, and the geometric
origin of the decoupling at hand is again the Hopf fibration of the
$\mathrm{AdS}_3$ in terms of an $H_2$.

\boldmath
\section{\boldmath $\mathrm{AdS}_2 \times S^2 $ \unboldmath}\label{ads2s2}
\unboldmath

The $\mathrm{AdS}_2 \times S^2$ geometry  appeared first in the
context of Reissner--Nordstr\"om black holes. The latter are
solutions of Maxwell--Einstein theory in four dimensions,
describing charged, spherically symmetric black holes. For a black
hole of mass $M$ and charge $Q$, the solution reads:
\begin{subequations}
  \begin{align}
    \di s^2 &= - \left( 1 - \frac{r_+}{r} \right)\left( 1 - \frac{r_-}{r}
    \right)\di t^2 + \frac{\di r^2}{\left( 1 - \frac{r_+}{r} \right)\left( 1
        - \frac{r_-}{r}
      \right)} + r^2 \di \Omega_{2}^2 \, ,\\
    F &= \frac{Q}{r^2} \ \di t \land \di r \hspace{1em}  \text{with}
    \hspace{1em} r_{\pm} = G_4 \left( M \pm \sqrt{M^2 - Q^2}
    \right);
  \end{align}
\end{subequations}
$r_+$ and $r_-$ are the outer and inner horizons, and $G_4$ is
Newton's constant in four dimensions.

In the extremal case, $r_+ = r_- = r_0$ ($M^2 = Q^2$), and the
metric approaches the $\mathrm{AdS}_2 \times S^2$ geometry in the
near-horizon\footnote{With the near-horizon coordinates
$U=(1-r_0/r)^{-1}$ and $T=t/r_0$, the near-horizon geometry is
$$\di s^2 = r_{0}^2 \left( - \frac{\di T^2 }{U^2} +\frac{\di U^2}{U^2} +  \di
\Omega_{2}^2\right).$$ Both $\mathrm{AdS}_2$ and $S^2$ factors
have the same radius $r_{0}$.} limit $r\to r_0$. This solution can
of course be embedded in various four-dimensional
compactifications of string theory, and will be supersymmetric in
the extremal case (see e.g.~\cite{Youm:1997hw} for a review). In
this paper we are dealing with some heterotic compactification.

In particular the $\mathrm{AdS}_2 \times S^2$ geometry appears 
in type IIB superstring theory, but with \textsc{rr}
backgrounds~\cite{Ferrara:1995ih}. The black hole solution is
obtained by wrapping D3-branes around 3-cycles of a Calabi--Yau
three-fold; in the extremal limit, one obtains the $\mathrm{AdS}_2
\times S^2$ solution, but at the same time the CY moduli freeze to
some particular values. A hybrid Green--Schwartz sigma-model
action for this model has been presented
in~\cite{Berkovits:1999zq} (see also~\cite{Verlinde:2004gt} for AdS$_2$).
The interest for $\mathrm{AdS}_2 \times S^2$ space--time is
motivated by the fact that  it provides an interesting candidate
for AdS/\textsc{cft} correspondence~\cite{Maldacena:1998re}. In
the present case the dual theory should correspond to some
superconformal quantum
mechanics~\cite{Boonstra:1998yu,Claus:1998ts,Gibbons:1998fa,Cadoni:2000gm}. 
According to some recent works~\cite{Ooguri:2004zv,Vafa:2004qa}, these 
string theory black holes are deeply related to topological strings 
on the Calabi-Yau manifold, leading to a prediction for their entropy.

\boldmath
\subsection{The spectrum}\label{ads2spec}
\unboldmath

As a first step in the computation of the $\mathrm{AdS}_2 \times
S^2$ string spectrum, we must determine the spectrum of the
$\mathrm{AdS}_2$ factor, by using the same limiting procedure as
in Sec.~\ref{pfs2} for the sphere. The spectrum of the
electrically deformed $\mathrm{AdS}_3$, is displayed in
Eqs.~(\ref{leftspecadsdef}) and (\ref{rightspecadsdef}). The
$\mathrm{AdS}_2$ limit is reached for $\cosh x \to \infty$, which
leads to the following constraint on the charges of the primary
fields:
\begin{equation}
\bar{n} + \frac{h}{2} + \sqrt{\frac{2}{k}} \left( \mu
+ n + \frac{a}{2} \right) = 0.
\label{ads2limit}
\end{equation}
In contrast with the $S^2$ case, since $\mu$ is any real number
--~irrespectively of the kind of $SL(2,\mathbb{R})$
representation~-- there is \emph{no extra} quantization
condition for the level to make this limit well-defined. In this
limit, the extra $U(1)$ decompactifies as usual and can be
removed. Plugging the constraint~(\ref{ads2limit}) in the
expressions for the dimensions of the affine primaries, we find
\begin{subequations}
  \label{primAdS2}
  \begin{align}
    L_0 &= -\frac{j(j-1)}{k} - \frac{1}{2} \left( \bar{n} + \frac{h}{2} \right)^2
    - \frac{1}{2} \left( n+ \frac{a}{2} \right)^2  , \label{AdS2L} \\
    \bar{L}_0 &= -\frac{j(j-1)}{k}.\label{AdS2R}
  \end{align}
\end{subequations}

In addition to the original $\mathrm{AdS}_3$ spectrum,
Eqs.~(\ref{leftspecads}) and (\ref{rightspecads}), the
right-moving part contain an extra fermionic lattice corresponding
to the states charged under the electric field. Despite the
absence of $N=2$ superconformal symmetry due to the Lorentzian
signature, the theory has a ``fermion-number'' left symmetry,
corresponding to the current:
\begin{equation}
J = \imath \psi^1 \psi^3 + \frac{2}{k} \left(J^2 + \imath \psi^1
\psi^3\right).
\end{equation}
The charges of the primaries (\ref{primAdS2}) are
\begin{equation}
\mathcal{Q}_F = n + \frac{a}{2} - \sqrt{\frac{2}{k}} \left( \bar{n} +
\frac{h}{2} \right).
\end{equation}

\boldmath
\subsection{$\mathrm{AdS}_2 \times S^2 \times \mathcal{M}$ and space--time
  supersymmetry} \unboldmath

Let us now consider the complete heterotic string background which
consists of the $\mathrm{AdS}_2 \times S^2$ space--time times an
$N=2$ internal conformal field theory $\mathcal{M}$, that we will
assume to be of central charge $\hat{c}=6$ and with integral
$R$-charges. Examples of thereof are toroidal or flat-space
compactifications, as well as Gepner models~\cite{Gepner:1988qi}.

The levels $k$ of $SU(2)$ and $\hat{k}$ of $SL(2,\mathbb{R})$ are
such that the string background is critical:
\begin{equation}
  \hat{c} = \frac{2(k-2)}{k} + \frac{2(\hat{k}+2)}{\hat{k}} =
  4 \implies k = \hat{k}.
\end{equation}
This translates into the equality of the radii of the
corresponding $S^2$ and $\mathrm{AdS}_2$ factors, which is in turn
necessary for supersymmetry. Furthermore, the charge quantization
condition for the two-sphere (Sec. \ref{gens2}) restricts further
the level to $k = 2p^2$, $p \in \mathbb{N}$.

In this system the total fermionic charge is
\begin{equation}
\mathcal{Q} = n + \frac{a}{2} - \frac{N-h/2}{p} + n' + \frac{a}{2}
- \frac{\bar{n}' + h/2}{p} + \mathcal{Q}_{\mathcal{M}}.
\end{equation}
Hence, assuming that the internal $N=2$ charge
$\mathcal{Q}_{\mathcal{M}}$ is integral, further constraints on
the electromagnetic charges of the theory are needed in order to
achieve space--time supersymmetry. Namely, we must only keep
states such that
\begin{equation}
  N + \bar{n'}  = 0 \mod p.
\end{equation}
This projection is some kind of generalization of Gepner models.
Usually, such a projection is supplemented in string theory by new
twisted sectors. We then  expect that, by adding on top of this
projection the usual GSO projection on odd fermion number, one
will obtain a space--time supersymmetric background. However, the
actual computation would need the knowledge of hyperbolic coset
characters of $SL(2,\mathbb{R})$ (i.e. Lorentzian black-hole
characters), and of their modular properties.  We can already
observe that this ``Gepner-like'' orbifold keeps only states which
are ``dyonic'' with respect to the electromagnetic field background.
Notice that, by switching other fluxes in the internal theory
$\mathcal{M}$ one can describe more general projections.

\section{Outlook}
\label{out}

The main motivation of this work was to analyze the landscape of the $S^3$
and $\mathrm{AdS}_3$ deformation. This analysis is performed from the
geometrical viewpoint with the symmetry as a guideline. The deformations
obtained in that way are then shown to be target spaces of exact marginal
perturbations of the $SU(2)_k$ and $SL(2,\mathbb{R})_k$ \textsc{wzw} models.

An important corollary of our analysis is that geometric cosets like $S^2 =
SU(2)/U(1)$ or $\mathrm{AdS}_2 = SL(2,\mathbb{R})/U(1)_{\rm space-like}$ can be
realized, with appropriate (chromo) magnetic or electric fields, as exact
conformal models, which hence provide new string backgrounds. They appear as
limiting asymmetric marginal deformations of \textsc{wzw} models, and are
therefore tractable conformal field theories leading to unitary strings.

\begin{figure}[htbp]
  \begin{center}
     \includegraphics[width=\textwidth]{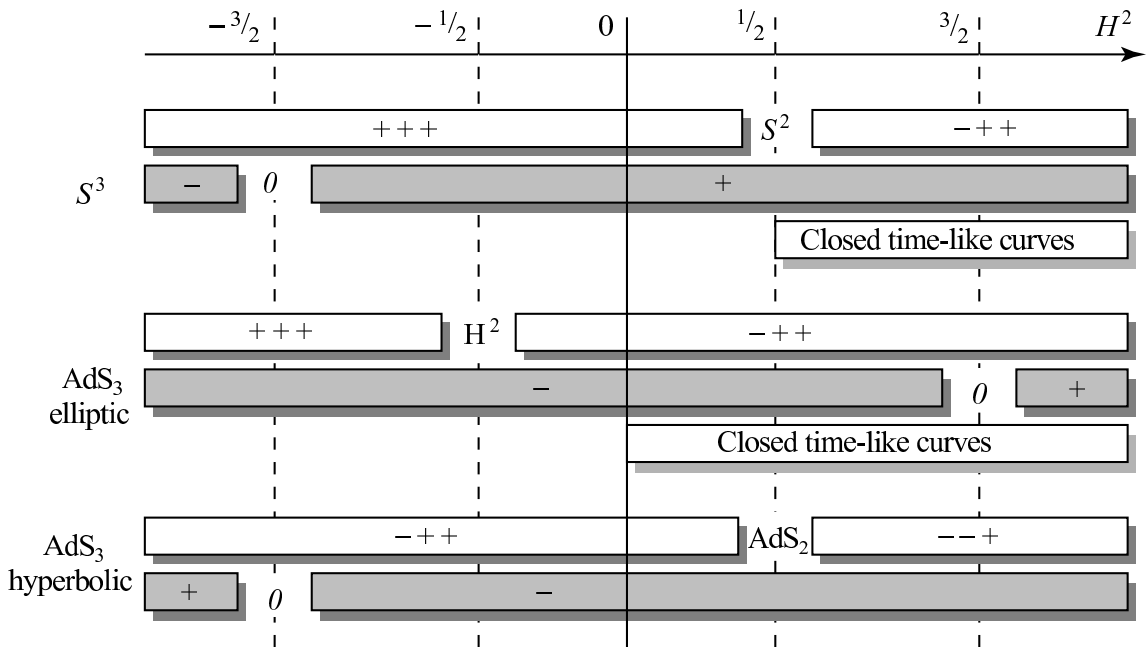}
  \end{center}
  \caption{We have summarized the various regimes appearing in
    magnetic/electric deformations of the $SU \left( 2\right)_k$ and $SL
    \left( 2, \setR \right)_k$ \textsc{wzw} models. Lighter bars indicate the
    signature, darker ones the sign of the curvature. Regions with $H^2<0$
    have well-defined geometries with imaginary gauge field. Their string
    interpretation is therefore questionable.}
  \label{fig:survey}
\end{figure}

The two-dimensional hyperbolic space $H_2= SL(2,\mathbb{R})/ U(1)_{\rm
  time-like}$ does also appear in the same manner, although the accompanying
magnetic field is imaginary. We display in Fig.~\ref{fig:survey} the summary
of the various situations analyzed here.

We have presented the spectrum and the partition function of the
$SU(2)_k$ magnetic deformation with $S^2 × \mathbb{R}$ as
extreme target space reached at $H^2_{\rm max} = 1/2$. At this
value the fiber $S^1$ decompactifies, as we have seen in Sec.
\ref{gens2}. We can go across and explore the deformation for
$H^2>1/2$. The geometry (metric given in Eq.~\eqref{S3metdef}) is
now Lorentzian and all other fields are real. This exact solution
is still a marginal deformation of $S^3$, of a peculiar type
though: it appears after a signature flip, has positive constant
curvature (see Fig. \ref{fig:survey}) and $U(1) × SU(2)$
isometry, where some of the $SU(2)$ Killing vectors are
time-like\footnote{In the regime $H^2>1/2$, $\d_\gamma$ is
time-like since $\| \d_\gamma \|^2 = -k\left( 2H^2 - 1\right)/4 $,
whereas  $\d_\alpha$ becomes time-like for $\cos^2 \beta >
1/2H^2$.}. In order to avoid trivial closed time-like curves, we
must promote the angle $\gamma$ to $\gamma \in \mathbb{R}$.
Unfortunately this is not enough to ensure consistency, and closed
time-like curves \emph{{à} la} Gödel appear whenever $\cos^2
\beta > 1/2H^2$, as one can check by using standard arguments
(Sec. \ref{ads}).

The space--time under consideration is a \emph{compact Gödel
universe} (``compact'' refers to the angle $\beta$), already
discussed in the context of general
relativity~\cite{Drukker:2003mg}. Our approach promotes it to the
level of exact string background, and we have in principle the
tools to investigate its spectrum. The latter may be obtained from
the partition function~(\ref{defspherespec}) with the replacement:
\begin{align}
  \cosh x = \frac{\imath}{2H^2 - 1} \ , &&  \sinh x = \imath
  \sqrt{\frac{2H^2}{2H^2 -1}}.
\end{align}
Such a straightforward approach is however questionable since it
involves a sort of analytic continuation. We will not expand further
on this issue.

The hyperbolic deformation of $SL(2,\mathbb{R})_k$, leading precisely to
$\mathrm{AdS}_2$, is an important achievement of this work. It is however technically
involved: no description is presently available for quantities like the
partition function, and our handling over the spectrum remains
incomplete\footnote{Much like in the compact Gödel universe quoted
  previously, one may try to give sense to the electric deformation for
  $H^2>1/2$. There, we have two times (see Fig.~\ref{fig:survey}). We can
  however perform an overall flip to reverse the signature $\set{--+} \to
  \set{++-}$ and obtain a well-defined geometry. As usual, the important
  issue is to implement all the sign flips and analytic continuations at the
  level of the \textsc{cft}.}. For the elliptic deformation, the partition function
is available in the ``Gödel regime'' ($H^2>0$) where the geometry is spoiled
by closed time-like curves. The continuation to the negative-$H^2$ region,
free of closed time-like curves, is not straightforward since it requires an
analytic continuation to imaginary magnetic field. The string theory is no
longer unitary, but it might still be useful to investigate this regime for
pure \textsc{cft} purposes.

Several other non-unitary regimes appear in the deformed geometries at hand,
either with Euclidean or with Lorentzian signature, and positive or negative
curvature (see Fig.~\ref{fig:survey}). Whether these could be connected to
the $H_3$ or $\mathrm{dS}_3$ spaces is an open question. Three-dimensional hyperbolic
plane, $H_3$, and de Sitter, $\mathrm{dS}_3$, are respectively negative and positive
constant-curvature symmetric spaces, with Euclidean and Lorentzian
signatures. The differences with respect to the $S^3$ or $\mathrm{AdS}_3$ are that
they have the opposite signature/curvature combination, and are not group
manifolds: $H_3$ and $\mathrm{dS}_3$ are $SL(2,\mathbb{C})$ cosets. A non-unitary
conformal sigma model can be defined on $H_3$, and considerable progress has
been made to understand its structure. In the case of $\mathrm{dS}_3$, nothing
similar is available, despite the growing interest for cosmological
applications. Any hint in that direction could be important.

New backgrounds arise by
introducing extra deformations. Since the residual affine symmetry of the
asymmetric marginal deformations is $U(1)_{\rm L}× SL(2,\mathbb{R})_{\rm
  R}$, there is room, at any point, for a extra modulus generated by a
$U(1)_{\rm L}× U(1)_{\rm R}\subset U(1)_{\rm L}× SL(2,\mathbb{R})_{\rm R}$
bilinear. This freedom has been used \emph{e.g} in~\cite{Israel:2003cx} in
order to cure the causal problem of the magnetic deformation 
(\emph{i.e.} the G\"odel universe), by slightly
deviating from the pure magnetic line. One can readily repeat this analysis
in the new regimes we have presented here, as already sketched for the
parabolic deformation -- end of Sec. \ref{pp} -- and further investigate the
$\mathrm{AdS}_3$ landscape and its connections to other three-dimensional
constant-curvature spaces of physical interest.

Our analysis deserves some extra comments. We are in position now
to describe $\mathrm{AdS}_2 × S^2$ as the target space of a
two-dimensional conformal model. This supersymmetric setting is
the near-horizon geometry of the extremal Reissner--Nordström
four-dimensional black hole. In our approach, this geometry arises
as a marginal deformation of the near-horizon geometry of the
\textsc{ns}5/F1 set up ($\mathrm{AdS}_3 × S^3$). The crucial question that
remains to be answered is how to deviate from this supersymmetric
geometry towards a non-extremal configuration of the charged
four-dimensional black hole.


We should also stress that the new string backgrounds we have constructed
may have interesting applications for holography. For example, the
AdS/plane-wave background (\ref{dsemdef}) is a superposition of two
solutions with known holographic interpretation. The line of geometries
relating $\mathrm{AdS}_3$ and $\mathrm{AdS}_2$ (hyperbolic deformation),
which are both candidates for AdS/\textsc{cft}, can possibly have some
holographic dual for all values of the modulus $H^2$. It is important to
remark that all the families of conformal field theories considered in this
paper preserve one chiral $SL(2,\mathbb{R})$ affine algebra, hence allow to
contruct one space--time Virasoro algebra by using the method given
in~\cite{Giveon:1998ns}.

Finally, concerning the general technique that we have used, it clearly
opens up new possibilities for compactifications.  Asymmetric deformations
can be performed on any group-$G$ manifold. The high residual symmetry $G×
U(1)^{{\rm rank}\ G}$ forces the geometry along the line as well as at the
boundary of the modulus.  There, we are naturally led to the geometric coset
$G/U(1)^{{\rm rank}\ G}$ times ${\rm rank}\ G$ decoupled, non compact, free
bosons. A magnetic field drives this asymmetric deformation in the case of
compact groups, whereas in non-compact cases a variety of situations can
arise as we showed for the $SL(2,\mathbb{R})$ \textsc{wzw} model.

\acknowledgments We have enjoyed very useful discussions with C.
Bachas, I. Bakas, E. Cremmer, B. Julia, E. Kiritsis and J. Troost.


\appendix

\section{Geometric deformations}\label{def}

In this appendix, we would like to present some selected results about
geometric deformations. Since most of the material discussed in this paper
deals with deformations of backgrounds that preserve part of the original
isometries, it is interesting to understand how such deformations can be
implemented in a controllable way.  Whether these are the most general, or
can be promoted to exact marginal lines in the underlying two-dimensional
sigma model, are more subtle issues that we analyze in the main part of the
paper for $S^3$ and $\mathrm{AdS}_3$.

Consider a manifold with metric $\di s^2 = g_{\alpha\beta} \di x^\alpha \di x^\beta$. We
assume the existence of an isometry generated by a Killing vector $\xi$. The
components of its dual form are $\varpi_\alpha = g_{\alpha\beta} \xi^\beta$. We will consider
the following family of geometries:
\begin{equation}
  g_{\alpha \beta}^{\rm def} = g_{\alpha \beta}^{\vphantom f} +
  f\left({\|\xi\|^2}\right) \varpi_\alpha^{\vphantom f}
  \varpi_\beta^{\vphantom f} \label{defmet}
\end{equation}
with $f$ an arbitrary function of the norm of the Killing vector.
Each member of the family corresponds to a specific choice of $f$.

Deformation (\ref{defmet}) provides a simple and straightforward recipe for
perturbing a background in a way that keeps under control its isometries.
Indeed, the vector $\xi$ is still an isometry generator (${\mathcal{L}}_\xi
\|\xi\|^2 =0; \; {\mathcal{L}}_\xi \varpi = 0$). So are all Killing vectors which
commute with $\xi$, whereas for a generic $f \left( \norm{\xi}^2 \right)$
those of the original isometry group which do not commute with $\xi$ are no
longer Killing vectors. Therefore, the ``deformed'' isometry group contains
only a subgroup of the simple component to which $\xi$ belongs, times the
other simple components, if any. These symmetry properties hold for generic
functions $f\left({\|\xi\|^2}\right)$. Accidental enhancements can occur
though, where a larger subgroup of the original group is restored (at least
locally).

The deformed dual form of $\xi$ is computed by
using (\ref{defmet}):
\begin{equation}
\varpi_\alpha^{\rm def} = \varpi_\alpha^{\vphantom f} \left\{ 1 +
\|\xi\|^2 f\left({\|\xi\|^2}\right)\right\} .\label{defform}
\end{equation}
Clearly, the symmetry requirement is not strong enough to reduce
substantially the freedom of the deformation. We would like to
focus on a specific subset of deformations for which
\begin{equation}
h = - \|\xi\|^2 f\left({\|\xi\|^2}\right) \label{defcon}
\end{equation}
is a real constant. This family is interesting for several
reasons. It is stable under repetition of the deformation: by
repeating this deformation we stay in the same class, but reach a
different point of it. This is a sort of integrability property
that leads to a \emph{one-parameter family} of continuous
deformations. The value $h=1$ is a \emph{critical} value of the
deformation parameter. At this value the metric degenerates. This
is a sign for the decoupling of one dimension. For $h>1$, the
signature flips and the decoupled dimension reenters the geometry,
with reversed signature though.

In the presence of Lorentzian geometries, light-like Killing
vectors can also be used. For those, Eq.~(\ref{defmet}) is
\emph{necessarily} of the form
\begin{equation}
g_{\alpha \beta}^{\rm def} = g_{\alpha \beta}^{\vphantom f} - h
\varpi_\alpha^{\vphantom f} \varpi_\beta^{\vphantom f}\ {\rm if}\
\ \|\xi\|^2=0.\label{defmetnul}
\end{equation}
The symmetry constraint is here very powerful. Furthermore, since
now $\varpi_\alpha^{\rm def} = \varpi_\alpha^{\vphantom f}$, no
critical phenomenon occurs in this case, at least for
\emph{finite} values of $h$.

Deformations of the kind (\ref{defmet}) with (\ref{defcon}) arise naturally
as background geometries of integrable marginal deformations of \textsc{wzw}
models. The integrability requirement for the \textsc{cft} deformation
selects the above one-parameter family of geometries. In turn, these
families exhibit limiting spaces.

It is useful to notice that deformations (\ref{defmet}) with
(\ref{defcon}) become trivial if $\xi$ is of the form $\xi = \d_1$
with $g_{\alpha\beta}$ block diagonal: $g_{\alpha 1} = g_{11}\,
\delta_{\alpha 1}$. In this case, $\varpi = g_{11} \di x^1$, and
$g_{\alpha\beta}$ is unaffected for $(\alpha,\beta)\neq (1,1)$,
whereas $g_{11}^{\rm def} = g_{11}^{\vphantom f}\left( 1-h\right)
$. The net effect of the deformation is a rescaling of the
coordinate $x^1$. The isometry group remains unaltered, at least
locally. In fact, if $x^1$ is an angle ($\xi $ is a compact
generator), the deformation amounts to introducing an angle
deficit, which brakes globally the original isometry group to the
subgroup described previously and introduces a conical
singularity.

We can illustrate the latter ``degenerate'' situation with a
simple example: the case of the two-dimensional Euclidean plane.
The isometry group has three generators: one rotation and two
translations. Consider the rotation and perform the deformation as
in (\ref{defmet}), (\ref{defcon}). The resulting manifold is a
cone. Although translation generators do not commute with the
rotation generator, translation invariance is still a symmetry,
locally -- since the only effect of the rotation is to rescale the
polar angle. Translation is however broken globally. The critical
value of the deformation ($h=1$) corresponds to the maximal angle
deficit, where the cone degenerates into a half-line.

Many deformations driven by $\xi_{(1)}, \xi_{(2)}, \ldots$ can be
performed simultaneously along the previous lines of thought. In
order to keep $\xi_{(1)}, \xi_{(2)}, \ldots$ in the isometry group
of the deformed metric, these must commute, hence must belong to
the Cartan subgroup of the original isometry group. The full
deformed group also contains those among the original Killing
vectors, which commute with the set $\{\xi_{(1)}, \xi_{(2)},
\ldots\}$.


Curvature tensors (Riemann, Ricci, Gauss) can be computed for the
deformed geometries under consideration (Eq.~(\ref{defmet})), in
terms of the original ones. How much they are altered depends on
the left-over symmetry and on the function $f$. We will not expand
further in this direction.

\boldmath
\section{The three-sphere}\label{sph}
\unboldmath
The commutation relations for the generators of $SU(2)$ are
\begin{align}
 \label{eq:comm-su2}
  \comm{ J^1 , J^2} = \imath J^3 && \comm{ J^2 , J^3} = \imath J^1 &&
  \comm{ J^3 , J^1 } = \imath J^2.
\end{align}
A two-dimensional realization is obtained by using the standard
Pauli matrices\footnote{The normalization of the generators with
respect to the Killing product in $\frak{su} \left( 2\right)$:
$\kappa \left( X, Y\right) = \tr \left( X Y\right)$ is such that
$\kappa \left( J^a, J^b \right) = 1/2$ and correspondingly the
root has length squared $\psi = 2$.}$\sigma^a$: $J^a =  \sigma^a /
2$.

The Euler-angle parameterization for $SU(2)$ is defined as:
\begin{equation}
g = {\rm e}^{\imath {\gamma \over 2} \sigma^3} {\rm e}^{\imath
{\beta \over 2} \sigma^1} {\rm e}^{\imath {\alpha \over 2}
\sigma^3}.
\end{equation}
The $SU(2)$ group manifold is a unit-radius
three-sphere. A three-sphere can be embedded in flat Euclidean
four-dimensional space with coordinates $(x^1,x^2,x^3,x^4)$, as
$(x^1)^2 + (x^2)^2 + (x^3)^2 + (x^4)^2 = L^2$. The corresponding
$SU(2)$ element $g$ is the following:
\begin{equation}
  g  =  L^{-1}\
  \begin{pmatrix}
    x^4 + \imath x^2 &  x^3 + \imath x^1 \\ -
      x^3 + \imath x^1 & x^4 - \imath x^2
  \end{pmatrix}. \label{4embS3}
\end{equation}

In general, the invariant metric of a group manifold can be
expressed in terms of the left-invariant Cartan--Maurer one-forms.
In the $SU(2)$ case under consideration (unit-radius $S^3$),
\begin{align}
  \mathcal{J}^1 =\frac{1}{2}\tr \left( \sigma^1 g^{-1} \di g \right),&&
  \mathcal{J}^2 = \frac{1}{2}\tr \left( \sigma^2 g^{-1} \di g
  \right),&& \mathcal{J}^3 = \frac{1}{2}\tr \left( \sigma^3 g^{-1} \di g
  \right)
\end{align}
and
\begin{equation}
  \label{eq:ads-metrix}
  \di s^2 = \sum_{i=1}^3 \mathcal{J}^i \otimes \mathcal{J}^i
\end{equation}
The volume form reads:
\begin{equation}
  \label{eq:ads-vf}
  \omega_{[3]} = \mathcal{J}^1 \land \mathcal{J}^2 \land
  \mathcal{J}^3.
\end{equation}

In the Euler-angle parameterization, Eq.~(\ref{eq:ads-metrix})
reads (for a radius-L three-sphere):
\begin{equation}
  \di s^2 =
  \frac{L^2}{4} \left( \di \alpha^2 + \di \gamma^2 + 2 \cos \beta
    \di \alpha \di \gamma + \di \beta^2\right),
\end{equation}
whereas (\ref{eq:ads-vf}) leads to
\begin{equation}
  \label{eq:su2-vf}
  \omega_{[3]} =\frac{L^3}{8} \sin \beta \di \alpha \land \di \beta \land \di
  \gamma.
\end{equation}
The Levi--Civita connection has scalar curvature $R = 6/L^2$.

The isometry group of the $SU(2)$ group manifold is generated by
left or right actions on $g$: $g\to hg$ or $g\to gh$ $\forall h
\in SU(2)$. From the four-dimensional point of view, it is
generated by the rotations $\zeta_{ab}= i\left( x_a\partial_b -
x_b
  \partial_a\right)$ with $x_a=\delta_{ab}x^b$. We list here explicitly the
six generators, as well as the group action they correspond to:
\begin{subequations}
  \label{eq:Killing-SU2}
  \begin{align}
    L_1 &= \frac{1}{2}\left(-\zeta_{32} + \zeta_{41}\right), & g &\to
    \mathrm{e}^{-\imath\frac{\lambda}{2}\sigma^1}g,
    \label{SL1} \\
    L_2 &= \frac{1}{2}\left(-\zeta_{43}-\zeta_{12} \right), &  g &\to
    \mathrm{e}^{\imath\frac{\lambda}{2}\sigma^2}g,
    \label{SL2} \\
    L_3 &= \frac{1}{2}\left(-\zeta_{31} - \zeta_{42}\right), & g&\to
    \mathrm{e}^{\imath\frac{\lambda}{2}\sigma^3}g,
    \label{SL3} \\
    R_1 &= \frac{1}{2}\left( \zeta_{41} + \zeta_{32}\right), & g&\to
    g\mathrm{e}^{\imath\frac{\lambda}{2}\sigma^1},
    \label{SR1} \\
    R_2 &= \frac{1}{2}\left(-\zeta_{43} + \zeta_{12}\right), &  g&\to
    g\mathrm{e}^{\imath\frac{\lambda}{2}\sigma^2},
    \label{SR2} \\
    R_3 &= \frac{1}{2}\left(\zeta_{31} - \zeta_{42}\right), & g&\to
    g\mathrm{e}^{\imath\frac{\lambda}{2}\sigma^3}.
    \label{SR3}
  \end{align}
\end{subequations}
Both sets satisfy the algebra (\ref{eq:comm-su2}). The norms
squared of the Killing vectors are all equal to $L^2/4$.

The currents of the $SU \left( 2 \right)_k $ \textsc{wzw} model
are easily obtained as:
\begin{align}
\label{eq:su2-currents}
  J^i = - k \tr \left(\imath \sigma^i \d g g^{-1} \right) &&
  \bar J^i = -k \tr \left(\imath \sigma^i  g^{-1} \db g
  \right),
\end{align}
where $L=\sqrt{k}$, at the classical level. Explicit expressions
are given in Tab. \ref{tab:su2-currents}.
\newcommand{\LSU}[0]{
  \begin{minipage}{.32\textwidth}
    \vspace{-.7em}
    \begin{small}
      \begin{gather*}
        \frac{\sin \gamma }{\sin \beta} \d_\alpha + \cos \gamma \d_\beta -
        \frac{\sin \gamma }{\tan \beta} \d_\gamma \\
        \frac{\cos \gamma }{\sin \beta} \d_\alpha - \sin \gamma \d_\beta -
        \frac{\cos \gamma }{\tan \beta} \d_\gamma \\
        \d_\gamma
      \end{gather*}
    \end{small}
    \vspace{-1.5em}
  \end{minipage}
}
\newcommand{\RSU}[0]{
  \begin{minipage}{.32\textwidth}
    \vspace{-.7em}
    \begin{small}
      \begin{gather*}
        - \frac{\sin \alpha }{\tan \beta } \d_\alpha + \cos \alpha \d_\beta
        + \frac{\sin \alpha }{\sin \beta} \d_\gamma \\
        \frac{\cos \alpha}{\tan \beta} \d_\alpha + \sin \alpha \d_\beta -
        \frac{\cos \alpha }{\sin \beta} \d_\gamma \\
        \d_\alpha
      \end{gather*}
    \end{small}
    \vspace{-1.5em}
  \end{minipage}
}
\newcommand{\JSU}[0]{
  \begin{minipage}{.3\textwidth}
    \vspace{-.7em}
    \begin{small}
      \begin{gather*}
         k \left( \sin \beta \sin \gamma \d \alpha + \cos
          \gamma \d b\right) \\[.5em]
         k \left( \cos \gamma \sin \beta \d \alpha - \sin \gamma \d \beta \right) \\[.5em]
         k \left(\d \gamma + \cos \beta \d \alpha \right)
      \end{gather*}
    \end{small}
    \vspace{-1.5em}
  \end{minipage}
}
\newcommand{\JbSU}[0]{
  \begin{minipage}{.3\textwidth}
    \vspace{-.7em}
    \begin{small}
      \begin{gather*}
         k \left( \cos \alpha \db \beta + \sin \alpha
          \sin \beta \db \gamma\right) \\[.5em]
         k \left( \sin \alpha \db \beta - \cos \alpha \sin \beta
          \db \gamma\right)\\[.5em]
         k \left( \db \alpha + \cos \beta \db \gamma\right)
      \end{gather*}
    \end{small}
    \vspace{-1.5em}
  \end{minipage}
}

\TABLE{\begin{tabular}{|c|c|c|}
    \hline sector & Killing vector& Current \\ \hline \hline
    \begin{rotate}{90}\hspace{-2em}left moving\end{rotate} & \LSU &
    \JSU \\ \hline
    \begin{rotate}{90}\hspace{-2.5em}right moving\end{rotate} & \RSU &
    \JbSU \\\hline
  \end{tabular}
  \caption{ Killing
  vectors $\set{\imath L_1,\imath L_2,\imath L_3}$ and $\set{\imath R_1,\imath R_2,\imath
  R_3}$, and holomorphic and anti-holomorphic currents (as defined in Eqs.
  (\ref{eq:Killing-SU2}) and (\ref{eq:su2-currents})) in Euler angles.}
  \label{tab:su2-currents}
}


\section{\boldmath $\mathrm{AdS}_3$\unboldmath}
\label{antids}

The commutation relations for the generators of the $SL(2,\mathbb{R})$
algebra are
\begin{align}
  \comm{ J^1 , J^2} = - \imath J^3 && \comm{ J^2 , J^3} = \imath J^1 &&
  \comm{ J^3 , J^1 } = \imath J^2.
  \label{comm}
\end{align}
The sign in the first relation is the only difference with respect
to the $SU(2)$ in Eq.~\eqref{eq:comm-su2}.

The three-dimensional anti-de-Sitter space is the universal
covering of the $SL(2,\mathbb{R})$ group manifold. The latter can
be embedded in a Lorentzian flat space with signature $(-,+,+,-)$
and coordinates $(x^0,x^1,x^2,x^3)$:
\begin{equation}
  g  =  L^{-1}\
  \begin{pmatrix}
    x^0 + x^2 & x^1 + x^3 \\ x^1 -
      x^3 & x^0 - x^2
  \end{pmatrix}, \label{4emb}
\end{equation}
where $L$ is the radius of $\mathrm{AdS}_3$. On can again introduce
Euler-like angles
\begin{equation} g = \mathrm{e}^{\imath (\tau+\varphi) \sigma_2/2}
\mathrm{e}^{\rho \sigma_1} \mathrm{e}^{\imath(\tau-\varphi)
\sigma_2/2} , \label{euler}
\end{equation}
which provide good global coordinates for $\mathrm{AdS}_3$ when $\tau\in
]-\infty,+\infty[$, $\rho\in [0,\infty[$, and $\varphi\in [0,2\pi]$.

An invariant metric (see Eq.~\eqref{eq:ads-metrix}) can be
introduced on $\mathrm{AdS}_3$. In Euler angles, the latter reads:
\begin{equation}
\di s^2= L^2\left[- \cosh ^2 \rho  \, \di \tau ^2 +\di \rho^2 +
\sinh^2 \rho \, \di \varphi^2\right]. \label{dseul}
\end{equation}
The Ricci scalar of the corresponding Levi--Civita connection is
$R=-6/L^2$.

The isometry group of the $SL(2,\mathbb{R})$ group manifold is
generated by left or right actions on $g$: $g\to hg$ or $g\to gh$
$\forall h \in SL(2,\mathbb{R})$. From the four-dimensional point
of view, it is generated by the Lorentz boosts or rotations
$\zeta_{ab}= i\left( x_a\partial_b - x_b
  \partial_a\right)$ with $x_a=\eta_{ab}x^b$. We list here explicitly the
six generators, as well as the group action they correspond to:
\begin{subequations}
  \label{eq:Killing-SL2R}
  \begin{align}
    L_1 &= \frac{1}{2}\left(\zeta_{32} - \zeta_{01}\right), & g &\to
    \mathrm{e}^{-\frac{\lambda}{2}\sigma^1}g,
    \label{L1} \\
    L_2 &= \frac{1}{2}\left(-\zeta_{31}-\zeta_{02} \right), &  g &\to
    \mathrm{e}^{-\frac{\lambda}{2}\sigma^3}g,
    \label{L2} \\
    L_3 &= \frac{1}{2}\left(\zeta_{03} - \zeta_{12}\right), & g&\to
    \mathrm{e}^{\imath\frac{\lambda}{2}\sigma^2}g,
    \label{L3} \\
    R_1 &= \frac{1}{2}\left( \zeta_{01} + \zeta_{32}\right), & g&\to
    g\mathrm{e}^{\frac{\lambda}{2}\sigma^1},
    \label{R1} \\
    R_2 &= \frac{1}{2}\left(\zeta_{31} - \zeta_{02}\right), &  g&\to
    g\mathrm{e}^{-\frac{\lambda}{2}\sigma^3},
    \label{R2} \\
    R_3 &= \frac{1}{2}\left(\zeta_{03} + \zeta_{12}\right), & g&\to
    g\mathrm{e}^{\imath\frac{\lambda}{2}\sigma^2}.
    \label{R3}
  \end{align}
\end{subequations}

Both sets satisfy the algebra (\ref{comm}). The norms of the Killing vectors
are the following:
\begin{equation}
  \norm{\imath L_1}^2 = \norm{ \imath R_1}^2 = \norm{\imath L_2}^2 =
  \norm{\imath R_2}^2 =- \norm{ \imath L_3}^2=-\norm{\imath R_3}^2 = \frac{L^2}{4}.
\end{equation}
Moreover $L_i \cdot L_j = 0$ for $i\neq j$ and similarly for the
right set. Left vectors are not orthogonal to right ones.

The isometries of the $SL(2,\mathbb{R})$ group manifold turn into symmetries
of the $SL(2,\mathbb{R})_k$ \textsc{wzw} model, where they are realized in
terms of conserved currents\footnote{When writing actions a choice of gauge
  for the \textsc{ns} potential is implicitly made, which breaks part of the
  symmetry: boundary terms appear in the transformations.  These must be
  properly taken into account in order to reach the conserved currents.
  Although the expressions for the latter are not unique, they can be put in
  an improved-Noether form, in which they have only holomorphic (for
  $L_i$'s) or anti-holomorphic (for $R_j$'s) components.}:
\begin{subequations}
  \label{eq:currents-SL2R}
\begin{align}
  J^1 \left( z \right) ± J^3 \left( z \right)&= -k \tr \left(
  \left(\sigma^1 \mp \imath \sigma^2\right)
  \d g
    g^{-1}\right), &
  J^2 \left( z \right) &= - k \tr \left( \sigma^3 \d g
    g^{-1}\right),
  \\
  \bar J^1 \left( \bar z \right) ± \bar J^3 \left( \bar z \right)&=
  k \tr \left( \left(\sigma^1 ± \imath \sigma^2\right) g^{-1}
    \db g \right), &
  \bar J^2 \left( \bar z \right) &= - k \tr \left( \sigma^3
    g^{-1}\db g \right).
\end{align}
\end{subequations}

At the quantum level, these currents, when properly normalized,
satisfy the following affine $SL(2,\mathbb{R})_k$
\textsc{opa}\footnote{In some conventions the level is $x=-k$. This
  allows to unify commutation relations for the affine
  $SL(2,\mathbb{R})_x$ and $SU(2)_x$ algebras. Unitarity demands
  $x<-2$ for the former and $0<x$ with integer $x$ for the latter.}:
\begin{subequations}
  \label{LOPA}
  \begin{align}
    J^3(z) J^3(0) & \sim - \frac{k}{2z^2},  \\
    J^3(z) J^{±}(0)& \sim ± \frac{J^{±}}{z},   \\
    J^+(z) J^-(0) & \sim \frac{2J^3}{z}-\frac{k}{z^2},
  \end{align}
\end{subequations}
and similarly for the right movers. The central charge of the enveloping
Virasoro algebra is $c= 3+ 6/(k-2)$.

We will introduce three different coordinate systems where the
structure of $\mathrm{AdS}_3 $ as a Hopf fibration is more transparent.
They are explicitly described in the following.
\begin{itemize}
\item The $\left( \rho, t, \phi \right)$ coordinate system used to describe 
the magnetic deformation in Sec.~\ref{mag} is defined as follows:
  \newcommand{\CR}[0]{\cosh \frac{\rho}{2}}
  \newcommand{\SR}[0]{\sinh \frac{\rho}{2}}
  \newcommand{\CPHI}[0]{\cosh \frac{\phi}{2}}
  \newcommand{\SPHI}[0]{\sinh \frac{\phi}{2}}
  \newcommand{\CT}[0]{\cos \frac{t}{2}}
  \newcommand{\ST}[0]{\sin \frac{t}{2}}
  \begin{equation}
    \begin{cases}
      \frac{x_0}{L} &= \CR \CPHI \CT - \SR \SPHI \ST \\
      \frac{x_1}{L} &= -\SR \SPHI \CT - \CR \SPHI \ST \\
      \frac{x_2}{L} &= -\CR \SPHI \CT + \SR \CPHI \ST \\
      \frac{x_3}{L} &= -\SR \SPHI \CT - \CR \CPHI \ST .
    \end{cases}
  \end{equation}
  The metric (\ref{eq:ads-metrix}) reads:
   \begin{equation}
     \label{eq:ads-rhotphi-metric}
     \di s^2 = \frac{L^2}{4} \left( \di \rho^2 + \di \phi^2 - \di t^2 -
       2 \sinh \rho \di t \di \phi\right)
   \end{equation}
   and the corresponding volume form is:
   \begin{equation}
     \label{eq:ads-rhotphi-vf}
     \omega_{[3]} = \frac{L^3}{8}\cosh \rho .
     \di \rho \land \di \phi  \land \di t
  \end{equation}
  Killing vectors and currents are given in Tab.
  \ref{tab:currents-timelike}. It is worth to remark that this coordinate
  system is such that the $t$-coordinate lines coincide with the integral
  curves of the Killing vector $\imath L_3$, whereas the $\phi$-lines are
  the curves of $\imath R_2$.
\item The $\left( r, x, \tau \right)$ coordinate system used to describe the
  electric deformation in Sec.~\ref{elec} is defined as follows:
  \renewcommand{\CR}[0]{\cosh \frac{r}{2}}
  \renewcommand{\SR}[0]{\sinh \frac{r}{2}}
  \newcommand{\CX}[0]{\cosh \frac{x}{2}}
  \newcommand{\SX}[0]{\sinh \frac{x}{2}}
  \renewcommand{\CT}[0]{\cos \frac{\tau}{2}}
  \renewcommand{\ST}[0]{\sin \frac{\tau}{2}}
  \begin{equation}
    \label{eq:ads-rxt-coo}
    \begin{cases}
      \frac{x_0}{L} &= \CR \CX \CT + \SR \SX \ST \\
      \frac{x_1}{L} &= - \SR \CX \CT + \CR \SX \ST \\
      \frac{x_2}{L} &= - \CR \SX \CT - \SR \CX \ST \\
      \frac{x_3}{L} &= \SR \SX \CT - \CR \CX \ST .
    \end{cases}
  \end{equation}
  For $\set{r,x,\tau} \in \mathbb{R}^3$, this patch covers exactly once the
  whole $\mathrm{AdS}_3$, and is regular
  everywhere~\cite{Coussaert:1994tu}.  The metric is then given by
  \begin{equation}
     \label{eq:ads-rxt-met}
    \di s^2 = \frac{L^2}{4} \left( \di r^2 + \di x^2 - \di \tau^2 +
      2 \sinh r \di x \di \tau \right)
  \end{equation}
  and correspondingly the volume form is
  \begin{equation}
    \label{eq:ads-rxt-vf}
    \omega_{[3]} = \frac{L^3}{8} \cosh r \di r \land \di x \land \di \tau .
  \end{equation}
  Killing vectors and currents are given in Tab.
  \ref{tab:currents-spacelike}. In this case the $x$-coordinate lines
  coincide with the integral curves of the Killing vector $\imath R_2$,
  whereas the $\tau$-lines are the curves of $\imath R_3$.
\item The Poincaré coordinate system used in Sec.~\ref{pp} to obtain the
  electromagnetic-wave background is defined by
  \begin{equation}
    \begin{cases}
    x^0 + x^2 &=\frac{L}{u}\\
    x^0 - x^2 &= Lu + \frac{L x^+ x^-}{u}\\
    x^1 ± x^3 &= \frac{L x^±}{u} .
    \end{cases}
  \end{equation}
  For $\set{u,x^+,x^-} \in \mathbb{R}^3$, the Poincaré
  coordinates cover once the $SL(2\mathbb{R})$ group manifold. Its
  universal covering, $\mathrm{AdS}_3$, requires an infinite
  number of such patches. Moreover, these coordinates exhibit
  a Rindler horizon at $\vert u \vert \to \infty$; the conformal
  boundary is at $\vert u \vert \to 0$.
  Now the metric reads:
  \begin{equation}
    \di s^2 = \frac{L^2}{u^2} \left( \di u^2 + \di x^+ \di x^- \right)
  \end{equation}
  and the volume form:
  \begin{equation}
    \label{eq:ads-poinc-volume}
    \omega_{[3]} = \frac{L^3}{2u^3} \di u \land \di x^- \land \di x^+ .
  \end{equation}
  In these coordinates it is simple to write certain a linear combination of
  the Killing vector so to obtain explicitly a light-like isometry
  generator. For this reason in Tab. \ref{tab:currents-poincare} we report
  the $\set{L_1+L_3, L_1-L_3, L_2, R_1+R_3, R_1-R_3, R_2 } $ isometry
  generators and the corresponding $\set{J_1+J_3, J_1-J_3, J_2, \bar J_1 +
    \bar J_3, \bar J_1 - \bar J_3, \bar J_2} $ currents.
\end{itemize}


\newcommand{\Lrhotphi}[0]{
  \begin{minipage}{.4\textwidth}
    \vspace{-.7em}
    \begin{small}
      \begin{gather*}
        \cos t \d_\rho + \frac{\sin t}{\cosh \rho} \d_\phi - \sin t \tanh
        \rho \d_t\\
        -\sin t \d_\rho + \frac{\cos t}{\cosh \rho} \d_\phi - \cos t \tanh
        \rho \d_t \\
        - \d_t
      \end{gather*}
    \end{small}
    \vspace{-1.5em}
  \end{minipage}
}
\newcommand{\Rrhotphi}[0]{
  \begin{minipage}{.4\textwidth}
    \vspace{-.7em}
    \begin{small}
      \begin{gather*}
        \cosh \phi \d_\rho - \sinh \phi \tanh \rho \d_\phi - \frac{\sinh
          \phi }{\cosh \rho} \d_t \\
        \d_\phi \\
        \sinh \phi \d_\rho - \cosh \phi \tanh \rho \d_\phi - \frac{\cosh
          \phi }{\cosh \rho} \d_t
      \end{gather*}
    \end{small}
    \vspace{-1.5em}
  \end{minipage}
}
\newcommand{\Jrhotphi}[0]{
  \begin{minipage}{.35\textwidth}
    \vspace{-.7em}
    \begin{small}
      \begin{gather*}
        k\left( \cos t \d \rho + \cosh \rho \sin t \d \phi
        \right)\\
        k \left( \cos t \cosh \rho \d \phi - \sin t \d \rho
        \right)\\
        k \left( \d t + \sinh \rho \d \phi\right)
      \end{gather*}
    \end{small}
    \vspace{-1.2em}
  \end{minipage}
}
\newcommand{\Jbrhotphi}[0]{
  \begin{minipage}{.35\textwidth}
    \vspace{-.7em}
    \begin{small}
      \begin{gather*}
        - k \left( \cosh \phi \db
          \rho + \cosh \rho \sinh \phi \db t \right)\\
        k \left( \db \phi - \sinh \rho \db
          t\right)\\
        k \left( \cosh \rho \cosh \phi \db t + \sinh \phi \db
          \rho \right)
      \end{gather*}
    \end{small}
    \vspace{-1.2em}
  \end{minipage}
}

\TABLE{\begin{tabular}{|c|c|c|}
    \hline sector & Killing vector& Current \\ \hline \hline
    \begin{rotate}{90}\hspace{-2em}left moving\end{rotate} & \Lrhotphi &
    \Jrhotphi\\ \hline
    \begin{rotate}{90}\hspace{-2.5em}right moving\end{rotate} & \Rrhotphi &
    \Jbrhotphi \\\hline
  \end{tabular}
  \caption{Killing
  vectors $\set{\imath L_1,\imath L_2,\imath L_3}$ and $\set{\imath R_1,\imath R_2,\imath
  R_3}$, and holomorphic and anti-holomorphic currents (as defined in Eqs.
  (\ref{eq:Killing-SL2R}) and (\ref{eq:currents-SL2R})) for
  the $ \left( \rho, t, \phi \right)$ coordinate system (elliptic base).}
  \label{tab:currents-timelike}
}


\newcommand{\Lrxtau}[0]{
  \begin{minipage}{.52\textwidth}
    \vspace{-.7em}
    \begin{small}
      \begin{gather*}
        \cosh x \d_r - \sinh x \tanh r \d_x + \frac{\sinh x}{\cosh r} \d_\tau \\
        \d_x \\
        - \sinh x \d_r + \cosh x \tanh r \d_x - \frac{\cosh x}{\cosh r}
        \d_\tau
      \end{gather*}
    \end{small}
    \vspace{-1.3em}
  \end{minipage}
}
\newcommand{\Rrxtau}[0]{
  \begin{minipage}{.52\textwidth}
    \vspace{-.7em}
    \begin{small}
      \begin{gather*}
        - \cos \tau \d_r + \frac{\sin \tau}{\cosh \tau } \d_x - \sin \tau
        \tanh \tau \d_\tau\\
        \frac{\left(\cos \tau + \sin \tau \tanh r\right) \d_x +\left( \cos
            \tau \sinh r - \frac{\sin \tau}{\cosh r}\right) \d_\tau}{\cosh
          r}\\
        - \d_\tau
      \end{gather*}
    \end{small}
    \vspace{-1.3em}
  \end{minipage}
}
\newcommand{\Jrxtau}[0]{
  \begin{minipage}{.32\textwidth}
    \vspace{-.7em}
    \begin{small}
      \begin{gather*}
        k \left( \cosh x \d r - \cosh r \sinh x \d \tau \right)\\
        k \left( \d x + \sinh r \d \tau \right)  \\
        k \left( \cosh r \cosh x \d \tau - \sinh x \d r \right)
      \end{gather*}
    \end{small}
    \vspace{-1.3em}
  \end{minipage}
}
\newcommand{\Jbrxtau}[0]{
  \begin{minipage}{.32\textwidth}
    \vspace{-.7em}
    \begin{small}
      \begin{gather*}
        k \left( -\cos \tau \db r + \cosh r
          \sin \tau \db r\right)    \\
        k \left( \cos \tau \cosh r \db x + \sin \tau
          \db r\right)    \\
        k \left( \db \tau - \sinh r \db x\right)
      \end{gather*}
    \end{small}
    \vspace{-1.3em}
  \end{minipage}
}

\TABLE{\begin{tabular}{|c|c|c|}
    \hline sector & Killing vector& Current \\ \hline \hline
    \begin{rotate}{90}\hspace{-2em}left moving\end{rotate} & \Lrxtau &
    \Jrxtau\\ \hline
    \begin{rotate}{90}\hspace{-2.5em}right moving\end{rotate} & \Rrxtau &
    \Jbrxtau \\\hline
  \end{tabular}
  \caption{Killing
  vectors $\set{\imath L_1,\imath L_2,\imath L_3}$ and $\set{\imath R_1,\imath R_2,\imath
  R_3}$, and holomorphic and anti-holomorphic currents (as defined in Eqs.
  (\ref{eq:Killing-SL2R}) and (\ref{eq:currents-SL2R})) for
  the $\left( r, x, \tau \right)$ coordinate system (hyperbolic base).}
  \label{tab:currents-spacelike}
}


\newcommand{\LPoin}[0]{
  \begin{minipage}{.3\textwidth}
    \vspace{-.7em}
    \begin{small}
      \begin{gather*}
        - \d_- \\[1em]
        u x^- \d_u -  u^2 \d_+ + \left( x^-\right)^2 \d_-\\[1em]
        \frac{u}{2} \d_u + x^- \d_-
      \end{gather*}
    \end{small}
    \vspace{-1.3em}
  \end{minipage}
}
\newcommand{\RPoin}[0]{
  \begin{minipage}{.3\textwidth}
    \vspace{-.7em}
    \begin{small}
      \begin{gather*}
        \d_+ \\[1em]
        - u x^+ \d_u - \left( x^+\right)^2 \d_+ + u^2 \d_- \\[1em]
         \frac{u}{2} \d_u +  x^+ \d_+
      \end{gather*}
    \end{small}
    \vspace{-1.3em}
  \end{minipage}
}
\newcommand{\JPoin}[0]{
  \begin{minipage}{.35\textwidth}
    \vspace{-.7em}
    \begin{small}
      \begin{gather*}
        -2k\frac{\partial x^+}{u^2} \\
         2k\left( 2 x^- \frac{\partial u}{u}-\partial x^- +
          (x^-)^2\frac{\partial x^+}{u^2} \right)\\
         2k\left( \frac{\partial u}{u} + x^- \frac{\partial
            x^+}{u^2}\right)
      \end{gather*}
    \end{small}
    \vspace{-1em}
  \end{minipage}
}
\newcommand{\JbPoin}[0]{
  \begin{minipage}{.35\textwidth}
    \vspace{-.7em}
    \begin{small}
      \begin{gather*}
         2k\frac{\bar\partial x^-}{u^2}\\
         2k\left(- 2 x^+ \frac{\bar\partial u}{u}+ \bar\partial
          x^+ - (x^+)^2\frac{\bar\partial x^-}{u^2} \right)\\
         2k\left( \frac{\bar\partial u}{u} + x^+
          \frac{\bar\partial x^-}{u^2}\right)
      \end{gather*}
    \end{small}
    \vspace{-1em}
  \end{minipage}
}

\TABLE{\begin{tabular}{|c|c|c|}
    \hline sector & Killing vector& Current \\ \hline \hline
    \begin{rotate}{90}\hspace{-2em}left moving\end{rotate} & \LPoin &
    \JPoin\\ \hline
    \begin{rotate}{90}\hspace{-2.5em}right moving\end{rotate} & \RPoin &
    \JbPoin \\\hline
  \end{tabular}
  \caption{Killing vectors, and holomorphic and anti-holomorphic currents (as defined
  in Eqs. (\ref{eq:Killing-SL2R}) and (\ref{eq:currents-SL2R})) in
    Poincaré coordinates (parabolic base). The $\set{\imath L_1+\imath L_3, \imath L_1-\imath L_3, \imath L_2,
    \imath R_1+\imath R_3, \imath R_1-\imath R_3,
      \imath  R_2 } $ isometry generators and the corresponding $\set{J_1+J_3,
      J_1-J_3, J_2, \bar J_1 + \bar J_3, \bar J_1 - \bar J_3, \bar J_2} $
    currents are represented so to explicitly obtain light-like isometry
    generators.}
  \label{tab:currents-poincare}
}

\section{Equations of motion}
\label{eom}

The general form for the marginal deformations we have studied is given by:
\begin{equation}
  \label{eq:general-deformation}
  S = \frac{1}{2\pi} \int \di^2 z \, \left( G^{\textsc{wzw}}_{\mu \nu } + B^{\textsc{wzw}}_{\mu \nu } \right)
  \partial x^{\mu} \bar \partial x^{\nu}
  + \frac{\sqrt{k k_G} H}{2\pi} \int \di^2 z \, J \bar J_G.
\end{equation}
It is not completely trivial to read off the deformed background
fields that correspond to this action. In this appendix we will
present a method involving a Kaluza--Klein reduction,
following~\cite{Horowitz:1995rf}. For simplicity we will consider
the bosonic string with vanishing dilaton. The right-moving gauge
current $\bar J_G$ used for the deformation has now a left-moving
partner and can hence be bosonized as $\bar J_G = \imath \bar
\partial \varphi $, $\varphi \left( z, \bar z \right) $ being
interpreted as an internal degree of freedom. The sigma-model
action is recast as
\begin{equation}
  \label{eq:KK-action}
  S = \frac{1}{2 \pi} \int \di^2 z \: \left( G_{\textsc{mn}} + B_{\textsc{mn}} \right)
  \partial x^{\textsc{m}} \bar \partial x^{\textsc{n}},
\end{equation}
where the $x^{\textsc{m}}, \textsc{m}=1,\ldots,4$ embrace the
group coordinates $x^\mu, \mu = 1,2,3$ and the internal $x^4
\equiv \varphi$:
\begin{equation}
  x^{\textsc{m}} = \left( \begin{tabular}{c}
      $  x^\mu $\\ \hline
      $ x^4$
    \end{tabular}\right).
\end{equation}
If we split accordingly the background fields, we obtain the following
decomposition:
\begin{align}
  G_{\textsc{mn}} = \left( \begin{tabular}{c|c}
      $G_{\mu\nu } $& $ G_{\varphi \varphi } A_\mu $\\ \hline
      $ G_{\varphi \varphi } A_\mu $& $G_{\varphi \varphi }$
    \end{tabular}\right), &&
  B_{\textsc{m}\textsc{n}} =  \left( \begin{tabular}{c|c}
      $B_{\mu\nu}$ & $B_{\mu 4}$ \\ \hline
      $-B_{\mu 4}$ & 0
    \end{tabular}\right),
\end{align}
and the action becomes:
\begin{multline}
  S = \frac{1}{2 \pi} \int \di z^2 \left\{ \left( G_{\mu \nu } + B_{\mu \nu} \right) \d x^\mu \db x^\nu
    + \left( G_{\varphi \varphi } A_\mu + B_{\mu 4} \right) \d x^\mu \db \varphi \right.\\
  \left. + \left( G_{\varphi \varphi } A_\mu - B_{\mu 4} \right)
    \d \varphi \db x^\mu + G_{\varphi \varphi } \d \varphi \db
    \varphi\right\}.
\end{multline}
  
We would like to put the previous expression in such a form that space--time
gauge invariance,
\begin{align}
  A_\mu & \to A_\mu + \d_\mu \lambda,  \\
  B_{\mu 4} & \to B_{\mu 4} + \d_\mu \eta,
\end{align}
is manifest. This is achieved as follows:
\begin{multline}
  S = \frac{1}{2 \pi } \int \di^2 z \: \left\{\left( \hat G_{\mu \nu } + B_{\mu
        \nu }\right) \d x^\mu \db x^\nu + B_{\mu 4} \left( \partial x^\mu \bar \partial \varphi
      - \partial \varphi \bar \partial x^\mu \right) \right. +\\ \left. + G_{\varphi \varphi } \left( \partial
      \varphi + A_\mu \partial x^\mu \right) \left( \bar \partial \varphi + A_\mu \bar \partial x^\mu
    \right)\right\},
\end{multline}
where $\hat G_{\mu \nu }$ is the Kaluza--Klein metric
\begin{equation}
  \hat G_{\mu \nu } = G_{\mu \nu } - G_{\varphi \varphi } A_{\mu } A_{\nu    }.
\end{equation}
Expression \eqref{eq:KK-action} coincides with
\eqref{eq:general-deformation} after the following
identifications:
\newcommand{\mJ}{\ensuremath{\mathcal{J}}}
\begin{subequations}
  \label{eq:KK-fields}
  \begin{align}
    \hat G_{\mu \nu } &=  \frac{k}{2} \left( \mJ_\mu \mJ_\nu - 2 \textsc{h}^2
      \tilde{\mJ}_\mu \tilde{\mJ}_\nu \right) \label{eq:KK-metric}\\
    B_{\mu \nu } &= \frac{k}{2} J_\mu \land J_\nu ,     \label{eq:B-field} \\
    B_{\mu 4} &= G_{\varphi \varphi } A_\mu  =  \textsc{h} \sqrt{\frac{k k_g}{2}} \tilde{\mJ}_\mu, \\
    A_{\mu} &=  \textsc{h} \sqrt{\frac{2k}{k_g}} \tilde{\mJ}_\mu, \label{eq:KK-em-field} \\
    G_{\varphi \varphi } & = \frac{k_g}{2}.
  \end{align}
\end{subequations}

The various backgrounds we have found  throughout this paper
correspond to truly marginal deformations of \textsc{wzw} models.
Thus, they are target spaces of exact \textsc{cft}'s. They solve,
however, the lowest-order (in $\alpha'$) equations since all
higher-order effects turn out to be captured in the shift $k \to k
+ 2$. These $\mathcal{O}\left(\alpha^\prime\right)$ equations are
obtained by using the bosonic action \eqref{eq:KK-action}, or
equivalently by writing the heterotic string equations of motion
for the fields in \eqref{eq:KK-fields}. They
read~\cite{Fradkin:1985ys}:
\begin{subequations}
  \label{beta}
    \begin{align}
      \delta c &= -R + \frac{k_G}{16} F_{\phantom{a}\mu \nu}
      {F}^{\mu\nu},\\
      {\beta^G}_{\mu \nu} &= R_{\mu \nu}^{\vphantom{n}} -
      \frac{1}{4} H_{\mu \rho \sigma}^{\vphantom{n}}
      H_{\nu}^{\phantom{\nu} \rho\sigma} - \frac{k_G}{4} F_{\mu \rho}^{\vphantom{n}}
      {F^a}_\mu^{\phantom{\mu}\rho} = 0, \\
      {\beta^B}_{\nu \rho} & = \nabla^\mu_{\vphantom{n}} H_{\mu \nu \rho}^{\vphantom{n}} = 0, \\
      {\beta^A}_\mu & = \nabla^\nu_{\vphantom{n}} F_{\mu \nu}^{\vphantom{n}} -
      \frac{1}{2} F^{\nu \rho}_{\vphantom{n}} H_{\mu \nu \rho}^{\vphantom{n}} =
      0,
    \end{align}
\end{subequations}
where
\begin{align}
  F_{\mu \nu } &= \d_\mu A_\nu - \d_\nu A_\mu, \\
  H_{\mu \nu \rho } &= \d_\mu B_{\nu \rho } - \frac{k_G}{4} A_\mu F_{\nu \rho } +
  \text{cyclic}.
\end{align}



\bibliography{Biblia}

\providecommand{\href}[2]{#2}\begingroup\raggedright\begin{thebibliography}{10}

\bibitem{Kounnas:1990ud}
C.~Kounnas, M.~Porrati, and B.~Rostand, {\it On {N=4} extended super{L}iouville
  theory},  {\em Phys. Lett.} {\bf B258} (1991) 61--69.

\bibitem{Callan:1991dj}
J.~Callan, Curtis~G., J.~A. Harvey, and A.~Strominger, {\it World sheet
  approach to heterotic instantons and solitons},  {\em Nucl. Phys.} {\bf B359}
  (1991) 611--634.

\bibitem{Antoniadis:1994sr}
I.~Antoniadis, S.~Ferrara, and C.~Kounnas, {\it Exact supersymmetric string
  solutions in curved gravitational backgrounds},  {\em Nucl. Phys.} {\bf B421}
  (1994) 343--372, [\href{http://xxx.lanl.gov/abs/hep-th/9402073}{{\tt
  hep-th/9402073}}].

\bibitem{Antoniadis:1990mn}
I.~Antoniadis, C.~Bachas, and A.~Sagnotti, {\it Gauged supergravity vacua in
  string theory},  {\em Phys. Lett.} {\bf B235} (1990) 255.

\bibitem{Boonstra:1998yu}
H.~J. Boonstra, B.~Peeters, and K.~Skenderis, {\it Brane intersections, anti-de
  {S}itter spacetimes and dual superconformal theories},  {\em Nucl. Phys.}
  {\bf B533} (1998) 127--162,
  [\href{http://xxx.lanl.gov/abs/hep-th/9803231}{{\tt hep-th/9803231}}].

\bibitem{Chaudhuri:1989qb}
S.~Chaudhuri and J.~A. Schwartz, {\it A criterion for integrably marginal
  operators},  {\em Phys. Lett.} {\bf B219} (1989) 291.

\bibitem{Forste:2003km}
S.~Forste and D.~Roggenkamp, {\it Current current deformations of conformal
  field theories, and wzw models},  {\em JHEP} {\bf 05} (2003) 071,
  [\href{http://xxx.lanl.gov/abs/hep-th/0304234}{{\tt hep-th/0304234}}].

\bibitem{Kiritsis:1995ta}
E.~Kiritsis and C.~Kounnas, {\it Infrared regularization of superstring theory
  and the one loop calculation of coupling constants},  {\em Nucl. Phys.} {\bf
  B442} (1995) 472--493, [\href{http://xxx.lanl.gov/abs/hep-th/9501020}{{\tt
  hep-th/9501020}}].

\bibitem{Kiritsis:1995iu}
E.~Kiritsis and C.~Kounnas, {\it Infrared behavior of closed superstrings in
  strong magnetic and gravitational fields},  {\em Nucl. Phys.} {\bf B456}
  (1995) 699--731, [\href{http://xxx.lanl.gov/abs/hep-th/9508078}{{\tt
  hep-th/9508078}}].

\bibitem{Israel:2003cx}
D.~Isra{ë}l, {\it Quantization of heterotic strings in a {G\"odel/anti de
  Sitter} spacetime and chronology protection},  {\em JHEP} {\bf 01} (2004)
  042, [\href{http://xxx.lanl.gov/abs/hep-th/0310158}{{\tt hep-th/0310158}}].

\bibitem{Giveon:1994ph}
A.~Giveon and E.~Kiritsis, {\it Axial vector duality as a gauge symmetry and
  topology change in string theory},  {\em Nucl. Phys.} {\bf B411} (1994)
  487--508, [\href{http://xxx.lanl.gov/abs/hep-th/9303016}{{\tt
  hep-th/9303016}}].

\bibitem{Johnson:1995kv}
C.~V. Johnson, {\it Heterotic coset models},  {\em Mod. Phys. Lett.} {\bf A10}
  (1995) 549--560, [\href{http://xxx.lanl.gov/abs/hep-th/9409062}{{\tt
  hep-th/9409062}}].

\bibitem{Lowe:1994gt}
D.~A. Lowe and A.~Strominger, {\it Exact four-dimensional dyonic black holes
  and bertotti- robinson space-times in string theory},  {\em Phys. Rev. Lett.}
  {\bf 73} (1994) 1468--1471,
  [\href{http://xxx.lanl.gov/abs/hep-th/9403186}{{\tt hep-th/9403186}}].

\bibitem{Tseytlin:1994my}
A.~A. Tseytlin, {\it Conformal sigma models corresponding to gauged
  {Wess-Zumino- Witten} theories},  {\em Nucl. Phys.} {\bf B411} (1994)
  509--558, [\href{http://xxx.lanl.gov/abs/hep-th/9302083}{{\tt
  hep-th/9302083}}].

\bibitem{Ferrara:1995ih}
S.~Ferrara, R.~Kallosh, and A.~Strominger, {\it {N=2} extremal black holes},
  {\em Phys. Rev.} {\bf D52} (1995) 5412--5416,
  [\href{http://xxx.lanl.gov/abs/hep-th/9508072}{{\tt hep-th/9508072}}].

\bibitem{Kutasov:1998zh}
D.~Kutasov, F.~Larsen, and R.~G. Leigh, {\it String theory in magnetic monopole
  backgrounds},  {\em Nucl. Phys.} {\bf B550} (1999) 183--213,
  [\href{http://xxx.lanl.gov/abs/hep-th/9812027}{{\tt hep-th/9812027}}].

\bibitem{Kiritsis:1988rv}
E.~Kiritsis, {\it Character formulae and the structure of the representations
  of the {N=1, N=2} superconformal algebras},  {\em Int. J. Mod. Phys.} {\bf
  A3} (1988) 1871.

\bibitem{Dobrev:1987hq}
V.~K. Dobrev, {\it Characters of the unitarizable highest weight modules over
  the {N=2} superconformal algebras},  {\em Phys. Lett.} {\bf B186} (1987) 43.

\bibitem{Matsuo:1987cj}
Y.~Matsuo, {\it Character formula of {$C < 1$} unitary representation of {N=2}
  superconformal algebra},  {\em Prog. Theor. Phys.} {\bf 77} (1987) 793.

\bibitem{Ravanini:1987yg}
F.~Ravanini and S.-K. Yang, {\it Modular invariance in {N=2} superconformal
  field theories},  {\em Phys. Lett.} {\bf B195} (1987) 202.

\bibitem{Berglund:1996dv}
P.~Berglund, C.~V. Johnson, S.~Kachru, and P.~Zaugg, {\it Heterotic coset
  models and (0,2) string vacua},  {\em Nucl. Phys.} {\bf B460} (1996)
  252--298, [\href{http://xxx.lanl.gov/abs/hep-th/9509170}{{\tt
  hep-th/9509170}}].

\bibitem{Forste:1994wp}
S.~Forste, {\it A truly marginal deformation of {SL(2, R)} in a null
  direction},  {\em Phys. Lett.} {\bf B338} (1994) 36--39,
  [\href{http://xxx.lanl.gov/abs/hep-th/9407198}{{\tt hep-th/9407198}}].

\bibitem{Israel:2003ry}
D.~Isra{ë}l, C.~Kounnas, and M.~P. Petropoulos, {\it Superstrings on {NS}5
  backgrounds, deformed {AdS(3)} and holography},  {\em JHEP} {\bf 10} (2003)
  028, [\href{http://xxx.lanl.gov/abs/hep-th/0306053}{{\tt hep-th/0306053}}].

\bibitem{Dijkgraaf:1992ba}
R.~Dijkgraaf, H.~Verlinde, and E.~Verlinde, {\it String propagation in a black
  hole geometry},  {\em Nucl. Phys.} {\bf B371} (1992) 269--314.

\bibitem{Rooman:1998xf}
M.~Rooman and P.~Spindel, {\it Goedel metric as a squashed anti-de sitter
  geometry},  {\em Class. Quant. Grav.} {\bf 15} (1998) 3241--3249,
  [\href{http://xxx.lanl.gov/abs/gr-qc/9804027}{{\tt gr-qc/9804027}}].

\bibitem{Reboucas:1983hn}
M.~J. Reboucas and J.~Tiomno, {\it On the homogeneity of riemannian space-times
  of {G\"odel} type},  {\em Phys. Rev.} {\bf D28} (1983) 1251--1264.

\bibitem{Drukker:2003sc}
N.~Drukker, B.~Fiol, and J.~Simon, {\it {G\"o}del's universe in a supertube
  shroud},  {\em Phys. Rev. Lett.} {\bf 91} (2003) 231601,
  [\href{http://xxx.lanl.gov/abs/hep-th/0306057}{{\tt hep-th/0306057}}].

\bibitem{Vilenkin}
N.~J. Vilenkin and A.~Klimyk, {\em Representation of {L}ie groups and special
  functions}.
\newblock Kluwer Academic Publishers, Dordrecht, 1991.

\bibitem{Maldacena:2000hw}
J.~M. Maldacena and H.~Ooguri, {\it Strings in {AdS(3) and SL(2,R) WZW model.
  I}},  {\em J. Math. Phys.} {\bf 42} (2001) 2929--2960,
  [\href{http://xxx.lanl.gov/abs/hep-th/0001053}{{\tt hep-th/0001053}}].

\bibitem{Petropoulos:1990fc}
P.~M.~S. Petropoulos, {\it Comments on {SU(1,1)} string theory},  {\em Phys.
  Lett.} {\bf B236} (1990) 151.

\bibitem{Aharony:1998ub}
O.~Aharony, M.~Berkooz, D.~Kutasov, and N.~Seiberg, {\it Linear dilatons,
  {NS}5-branes and holography},  {\em JHEP} {\bf 10} (1998) 004,
  [\href{http://xxx.lanl.gov/abs/hep-th/9808149}{{\tt hep-th/9808149}}].

\bibitem{Banados:1992wn}
M.~Banados, C.~Teitelboim, and J.~Zanelli, {\it The black hole in
  three-dimensional space-time},  {\em Phys. Rev. Lett.} {\bf 69} (1992)
  1849--1851, [\href{http://xxx.lanl.gov/abs/hep-th/9204099}{{\tt
  hep-th/9204099}}].

\bibitem{Banados:1993gq}
M.~Banados, M.~Henneaux, C.~Teitelboim, and J.~Zanelli, {\it Geometry of the
  (2+1) black hole},  {\em Phys. Rev.} {\bf D48} (1993) 1506--1525,
  [\href{http://xxx.lanl.gov/abs/gr-qc/9302012}{{\tt gr-qc/9302012}}].

\bibitem{Natsuume:1998ij}
M.~Natsuume and Y.~Satoh, {\it String theory on three dimensional black holes},
   {\em Int. J. Mod. Phys.} {\bf A13} (1998) 1229--1262,
  [\href{http://xxx.lanl.gov/abs/hep-th/9611041}{{\tt hep-th/9611041}}].

\bibitem{Hemming:2001we}
S.~Hemming and E.~Keski-Vakkuri, {\it The spectrum of strings on {BTZ} black
  holes and spectral flow in the {SL(2,R) WZW} model},  {\em Nucl. Phys.} {\bf
  B626} (2002) 363--376, [\href{http://xxx.lanl.gov/abs/hep-th/0110252}{{\tt
  hep-th/0110252}}].

\bibitem{Coussaert:1994tu}
O.~Coussaert and M.~Henneaux, {\it Self-dual solutions of 2+1 {E}instein
  gravity with a negative cosmological constant},
  \href{http://xxx.lanl.gov/abs/hep-th/9407181}{{\tt hep-th/9407181}}.

\bibitem{Gawedzki:1991yu}
K.~Gawedzki, {\it Noncompact {WZW} conformal field theories},
  \href{http://xxx.lanl.gov/abs/hep-th/9110076}{{\tt hep-th/9110076}}.

\bibitem{Teschner:1997ft}
J.~Teschner, {\it On structure constants and fusion rules in the
  {SL(2,C)/SU(2)} wznw model},  {\em Nucl. Phys.} {\bf B546} (1999) 390--422,
  [\href{http://xxx.lanl.gov/abs/hep-th/9712256}{{\tt hep-th/9712256}}].

\bibitem{Youm:1997hw}
D.~Youm, {\it Black holes and solitons in string theory},  {\em Phys. Rept.}
  {\bf 316} (1999) 1--232, [\href{http://xxx.lanl.gov/abs/hep-th/9710046}{{\tt
  hep-th/9710046}}].

\bibitem{Berkovits:1999zq}
N.~Berkovits, M.~Bershadsky, T.~Hauer, S.~Zhukov, and B.~Zwiebach, {\it
  Superstring theory on {AdS(2) $\times$ S(2)} as a coset supermanifold},  {\em
  Nucl. Phys.} {\bf B567} (2000) 61--86,
  [\href{http://xxx.lanl.gov/abs/hep-th/9907200}{{\tt hep-th/9907200}}].

\bibitem{Verlinde:2004gt}
H.~Verlinde, {\it Superstrings on {AdS(2)} and superconformal matrix quantum
  mechanics},  \href{http://xxx.lanl.gov/abs/hep-th/0403024}{{\tt
  hep-th/0403024}}.

\bibitem{Maldacena:1998re}
J.~M. Maldacena, {\it The large {N} limit of superconformal field theories and
  supergravity},  {\em Adv. Theor. Math. Phys.} {\bf 2} (1998) 231--252,
  [\href{http://xxx.lanl.gov/abs/hep-th/9711200}{{\tt hep-th/9711200}}].

\bibitem{Claus:1998ts}
P.~Claus {\em et.~al.}, {\it Black holes and superconformal mechanics},  {\em
  Phys. Rev. Lett.} {\bf 81} (1998) 4553--4556,
  [\href{http://xxx.lanl.gov/abs/hep-th/9804177}{{\tt hep-th/9804177}}].

\bibitem{Gibbons:1998fa}
G.~W. Gibbons and P.~K. Townsend, {\it Black holes and {C}alogero models},
  {\em Phys. Lett.} {\bf B454} (1999) 187--192,
  [\href{http://xxx.lanl.gov/abs/hep-th/9812034}{{\tt hep-th/9812034}}].

\bibitem{Cadoni:2000gm}
M.~Cadoni, P.~Carta, D.~Klemm, and S.~Mignemi, {\it {AdS(2)} gravity as
  conformally invariant mechanical system},  {\em Phys. Rev.} {\bf D63} (2001)
  125021, [\href{http://xxx.lanl.gov/abs/hep-th/0009185}{{\tt
  hep-th/0009185}}].

\bibitem{Ooguri:2004zv}
H.~Ooguri, A.~Strominger, and C.~Vafa, {\it Black hole attractors and the
  topological string},  \href{http://xxx.lanl.gov/abs/hep-th/0405146}{{\tt
  hep-th/0405146}}.

\bibitem{Vafa:2004qa}
C.~Vafa, {\it Two dimensional yang-mills, black holes and topological strings},
   \href{http://xxx.lanl.gov/abs/hep-th/0406058}{{\tt hep-th/0406058}}.

\bibitem{Gepner:1988qi}
D.~Gepner, {\it Space-time supersymmetry in compactified string theory and
  superconformal models},  {\em Nucl. Phys.} {\bf B296} (1988) 757.

\bibitem{Drukker:2003mg}
N.~Drukker, B.~Fiol, and J.~Simon, {\it Goedel-type universes and the {L}andau
  problem},  \href{http://xxx.lanl.gov/abs/hep-th/0309199}{{\tt
  hep-th/0309199}}.

\bibitem{Giveon:1998ns}
A.~Giveon, D.~Kutasov, and N.~Seiberg, {\it Comments on string theory on
  {AdS(3)}},  {\em Adv. Theor. Math. Phys.} {\bf 2} (1998) 733--780,
  [\href{http://xxx.lanl.gov/abs/hep-th/9806194}{{\tt hep-th/9806194}}].

\bibitem{Horowitz:1995rf}
G.~T. Horowitz and A.~A. Tseytlin, {\it A new class of exact solutions in
  string theory},  {\em Phys. Rev.} {\bf D51} (1995) 2896--2917,
  [\href{http://xxx.lanl.gov/abs/hep-th/9409021}{{\tt hep-th/9409021}}].

\bibitem{Fradkin:1985ys}
E.~S. Fradkin and A.~A. Tseytlin, {\it Quantum string theory effective action},
   {\em Nucl. Phys.} {\bf B261} (1985) 1--27.

\end{thebibliography}\endgroup

\end{document}